\documentclass[showpacs,aps,prd,10pt,nofootinbib,twocolumn]{revtex4-1}
   
\usepackage{graphicx}
\usepackage{psfrag}
\usepackage{epsf}
\usepackage[colorlinks=true,linkcolor=blue,citecolor=blue,urlcolor=blue]{hyperref}
\usepackage{amsmath}
\usepackage{amsfonts}
\usepackage{csquotes}

\usepackage{color}

\providecommand{\abs}[1]{\left|#1\right|}
\providecommand{\ep}[1]{{e}^{#1}}

\providecommand{\WKB}[0]{\textsc{WKB }}

\newcommand{\be}{\begin{equation}}
\newcommand{\ee}{\end{equation}}

\newcommand{\om}{\omega}
\newcommand{\Om}{\Omega}

\newcommand{\eq}[1]{Eq.~\eqref{#1}}

\begin{document}

\title{Dispersive fields in de Sitter space and event horizon thermodynamics}

\author{Xavier Busch}
\email{xavier.busch@th.u-psud.fr}
\affiliation{Laboratoire de Physique Th\'eorique, CNRS UMR 8627, B{\^{a}}t. 210, 
\\ Universit\'e Paris-Sud 11, 91405 Orsay Cedex, France}
\author{Renaud Parentani}
\email{renaud.parentani@th.u-psud.fr}
\affiliation{Laboratoire de Physique Th\'eorique, CNRS UMR 8627, B{\^{a}}t. 210, 
\\ Universit\'e Paris-Sud 11, 91405 Orsay Cedex, France}


\begin{abstract}
When Lorentz invariance is violated at high energy, the laws of black hole thermodynamics are apparently no longer satisfied. To shed light on this observation, we study dispersive fields in de~Sitter space. We show that the Bunch-Davies vacuum state restricted to the static patch is no longer thermal, and that the Tolman law is violated. However we also show that, for free fields at least, this vacuum is the only stationary stable state, as if it were in equilibrium. We then present a precise correspondence between dispersive effects found in de Sitter and in black hole metrics. This indicates that the consequences of dispersion on thermodynamical laws could also be similar. 
\end{abstract}

\date{\today}
\maketitle

\section{Introduction}

\vspace{-0.45cm}
The two main predictions of quantum field theory in curved space, namely black hole radiation~\cite{Hawking:1974sw,birrell1984quantum,Brout:1995rd} and primordial spectra in inflation~\cite{Starobinsky:1982ee,Mukhanov:1982nu,Mukhanov:1990me} share many properties. In particular, both spectra stem from vacuum fluctuations with extremely short wave lengths~\cite{Jacobson:1999zk}. They are therefore in principle sensitive to the ultrahigh frequency behavior of the theory. To check this sensitivity, following~\cite{Unruh:1994je}, nonlinear dispersion relations, which break the local Lorentz invariance, have been used in the context of black holes~\cite{Brout:1995wp,Corley:1996ar,Balbinot:2006ua,Unruh:2004zk,Coutant:2011in} and in cosmology~\cite{Martin:2000xs,Niemeyer:2000eh,Niemeyer:2001qe,Macher:2008yq}. However so far, these studies have been conducted separately and with different means. In homogeneous cosmology, time dependent modes with a fixed comoving wave vector have been used, whereas for black holes, the analysis was based on stationary modes. In spite of this, the two cases are unexpectedly similar, as we shall show.

In the present work, we analyze dispersive fields in de Sitter space for two reasons. First, since de Sitter endowed with a cosmological preferred frame is both homogeneous and stationary, high frequency dispersion can be studied along both approaches. This will allow us to relate them in a very precise way. We shall see that their compatibility relies on a two-dimensional symmetry group which is a subgroup of the de Sitter isometry group~\cite{Eling:2006xg}. Because the generators of the two symmetries do not commute, in each approach only one symmetry is manifest, while the other is somehow hidden. In fact, this extra symmetry has been exploited in the black hole near horizon approximation of Refs.~\cite{Brout:1995wp,Corley:1996ar,Balbinot:2006ua,Unruh:2004zk,Coutant:2011in}, but without noticing (in general) that it relies on properties that are exact in de Sitter space.

Second, the main consequence of high frequency dispersion, that is the loss of the thermality of the spectrum, has raised deep questions concerning the relationships between Lorentz symmetry and black hole thermodynamics~\cite{Jacobson:1991gr,Jacobson:1993hn,Jacobson:1996zs}. It has been claimed that this loss should lead to violations of the second law~\cite{Dubovsky:2006vk,Eling:2007qd,Jacobson:2008yc}. These issues are particularly relevant when working with extended theories of gravity, such as Einstein-aether~\cite{Eling:2004dk} or Horava gravity~\cite{Horava:2009uw}, see Ref.~\cite{Blas:2011ni}. To consider them in a simpler context, we study the Bunch-Davies (BD) vacuum of dispersive fields propagating in de Sitter space. In practice, we analyze the two-point function evaluated in this state. For all dispersion relations, we show that it is stationary and periodic in imaginary time with period $2\pi /H$, where $H$ is the Hubble factor, as it is for Lorentz invariant theories. However, in spite of this, we demonstrate that the BD vacuum is no longer a thermal state when restricted to the static patch (because the two-point function looses an analytical property which is part of the KMS conditions~\cite{Kay:1988mu,Brout:1995rd}). We also argue that the violations of thermality find their origin in the fact that the stationary Hamiltonian restricted to the static patch is no longer bounded from below, and that this loss is {\it not} necessarily related to the presence of event horizons. (For subluminal dispersion it can occur in stationary backgrounds without Killing horizon.) In spite of these violations, for free fields, we show that the BD vacuum is the only stationary state which is regular (in a sense that shall be made precise in the text) and that the other regular states \enquote{flow} towards this state. We expect that this will remain true for interacting theories, such as $\lambda \phi^4$. We then argue that the lessons obtained in de Sitter should apply to black holes because there is a precise correspondence between dispersive effects in de Sitter and in black hole metrics.

The paper is organized as follows. In Sec.~\ref{setting} we present the basic properties of high frequency dispersion in de Sitter space. In Sec.~\ref{thermal}, we demonstrate that the Bunch-Davies vacuum is no longer thermal for any (superluminal) dispersion relation. We also show that it is the only stationary stable state. In Sec.~\ref{pfour}, the departures from thermality and the $S$-matrix are exactly calculated for a quartic superluminal dispersion relation. We summarize our results in the conclusions section. In Appendix~\ref{group}, we discuss the group theoretical properties characterizing high frequency dispersion in de Sitter, and in Appendix~\ref{BHdSc} we study the correspondence between de Sitter and black holes.

\section{Dispersive fields in de Sitter space} 
\label{setting}

\subsection{Ultraviolet dispersion}  
\label{dSspace}

We work in 1+1 dimensions and consider the flat sections of de Sitter space. They can be described by $ds^2 = -dt^2 + a^2 dz^2$, where $a = {\ep{H t}}/{H} $ is the scale factor, and $t$ the cosmological time orthogonal to the flat sections. We assume that the preferred frame associated with high energy dispersion coincides with the cosmological frame. Following Ref.~\cite{Jacobson:1996zs}, we describe it by a unit timelike vector field $u$, treated as a given background field. When considering its dynamics it can be shown that $u$ flows to the cosmological rest frame~\cite{Jacobson:2000xp,Lim:2004js,Li:2007vz}, like peculiar velocities in expanding universes flow to rest. We also introduce the unit spacelike vector $s$ that is orthogonal to $u$. We use covariant expressions because we want to be able to transpose the present description to black hole metrics. In the above coordinate system $(t,z)$, one has $u^\mu = (-1,0)$, and $s^\mu = (0,{1}/{a})$. 

The fields $u,\, s$ define the preferred frequency and momentum by $\Omega \doteq u^\mu \partial_\mu , \, P  \doteq s^\mu \partial_\mu $. The dispersion relation then reads
\be
\label{disprel}
\Omega^2 = F^2(P^2) = m^2 + P^2 + f(P^2) .
\ee
We suppose that $f$ vanishes faster than $P^2$ for $P \to 0$, so as to recover a relativistic relation for $ P \ll \Lambda$, where $\Lambda$ gives the ultraviolet dispersive scale. As such, Eq.~\eqref{disprel} can be viewed as the Hamilton-Jacobi equation for the corresponding dispersive particle~\cite{Brout:1995wp,Balbinot:2006ua}. Using $g_{\mu\nu} + u_\mu u_\nu = s_\mu s_\nu $, this equation reads 
\be
g^{\mu \nu} \partial_\mu S \partial_\nu S  +m^2 + f\!\left( (s^\mu \partial_\mu S )^2 \right) = 0 ,
\label{dispHJeq}
\ee
where $S(t,z)$ is the action of the particle. On the other hand, Eq.~\eqref{disprel} can also be viewed as the dispersion relation governing some field. However there is some ambiguity because of the ordering of the differential operators, and nonminimal couplings. In this paper, we work with~\cite{Jacobson:2000gw,Lemoine:2001ar} 
\be
\left[ -  g^{\mu \nu} D_\mu D_\nu  +m^2 + f( - s^\mu D_\mu  s^\nu D_\nu ) \right] \Phi = 0 , 
\label{dispKGeq}
\ee
where $D_\mu$ is the covariant derivative. For other approaches based on condensed matter models, see Refs.~\cite{Schutzhold:2002rf,Macher:2009nz,Unruh:2012ve}.

\begin{figure}[!ht] 
\centering
\includegraphics[trim = 0cm 17cm 0cm 0cm, width=8cm]{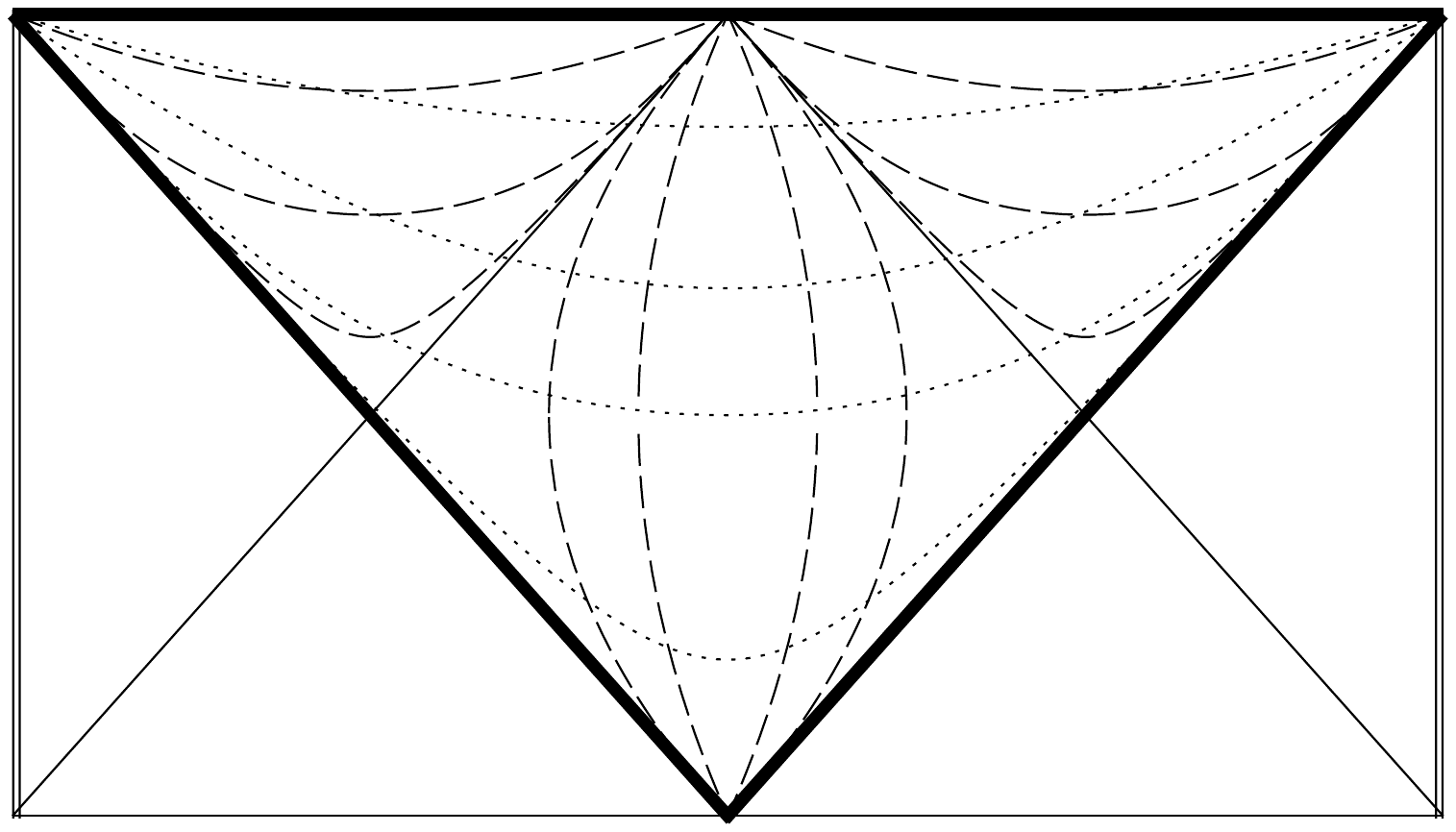} 
\caption{The Penrose diagram of the two-dimensional de Sitter space. The triangle surrounded by the thick line
characterizes the Poincar\'e patch covered by the coordinates $t,X$ of \eq{PG}. Dotted lines represent $t=$ Cst, and dashed lines $X=$ Cst. The static patch $-1<HX<1$ is the square of the middle of the figure.} 
\label{penrose}
\end{figure}

At this point it should be observed that the homogeneous Killing field $K_z= \partial_z$ commutes with our $u$ field. In cosmological backgrounds, the comoving momentum $k= \partial_z = aP $ is thus conserved for all dispersion relations $f$. What is peculiar about de Sitter space is that the settings are also stationary. This can be easily seen when using the preferred coordinates $(t, X)$ defined by $ dt \doteq u_\mu dx^\mu$ and $\partial_X \doteq s^\mu \partial_\mu$, see Fig.~\ref{penrose}. (They are often called Lema\^{\i}tre or Painlev\'e-Gullstrand coordinates~\cite{Barcelo:2005fc}.) 
 The preferred time coincides with the cosmological time, and the new spatial coordinate is related to $z$ by $X = a(t)\,  z$. In these coordinates, the de Sitter metric reads 
\begin{equation}
\begin{split}
ds^2& = -dt^2 + (dX - v(X) \, dt )^2,
\end{split}
\label{PG}
\end{equation}
where $v = HX$. The vector field $K_t= - \partial_t\vert_X$, with the derivative defined at fixed $X$, is manifestly a stationary Killing field. One also verifies that $K_t$ commutes with $u$. In the presence of dispersion, this is a {necessary} condition for its eigenvalue $\om = -  \partial_t\vert_X$ to be conserved. The relation between the preferred and the constant frequency is $\Om = \om - v(X) P$.

Because of these two symmetries, Eqs.~\eqref{dispHJeq} and~\eqref{dispKGeq} can be analyzed either in the $k,t$ representation as done in cosmology~\cite{Martin:2000xs,Niemeyer:2000eh,Niemeyer:2001qe,Macher:2008yq}, or in the $\om,X$ representation as done in black hole physics~\cite{Brout:1995wp,Corley:1996ar,Balbinot:2006ua}. However one cannot simultaneously exploit both symmetries because the two Killing fields do not commute. In fact they obey~\cite{Eling:2006xg} 
\be
[K_t,K_z] = H K_z. 
\label{alg}
\ee
This is the algebra of a subgroup of the two-dimensional de Sitter group $SO(1,2)$. It corresponds to the affine group of $\mathbb{R}$: the algebra contains the translation operator $\partial_z$ and the dilatation operator $H z \partial_z$ acting on functions of $z$. Unlike the full de Sitter group, it is compatible with dispersion and/or dissipation. As explained in Appendix~\ref{group}, this follows from the fact that both $\Omega$ and $P$ are invariant under its action.

The connection with black hole physics is easily made since the norm of $K_t$ is $K_t^2 = - 1 +(HX )^2$. It vanishes on the (black) horizons located at $HX_\pm = \pm 1 $. We call them black because the preferred momentum at fixed $\omega$ decreases as $P \sim \ep{-Ht}$, as found near the horizon in black hole metrics endowed with a freely falling frame~\cite{Brout:1995wp,Balbinot:2006ua}. In fact, the differences between that situation and the present one only arise from the velocity profile $v$. In de Sitter, $v=HX$ is globally linear; whereas, for black holes, $v$ is linear near the horizon region only in a finite domain. As a result, if the stationary Killing field $K_t= - \partial_t\vert_X$ is common to both cases, $K_z$ is only an \enquote{approximate} Killing field in black hole backgrounds, as explained in Appendix~\ref{BHdSc}.

\subsection{Mode analysis} 
\label{Mode analysis}

Given the two symmetries, the solutions of  Eq.~\eqref{dispKGeq} can be decomposed either as 
\be
\Phi = \int_{-\infty}^{\infty} \!\!\frac{d{\bf{k}}}{\sqrt{2\pi}} \ep{i {\bf{k}}z}  \phi_{\bf{k}}(t), 
\label{kdec}
\ee
or as
\be
\Phi = \int_{-\infty} ^{\infty}\!\! \frac{ d\omega}{\sqrt{2\pi }}  \ep{- i \om t}  \phi_\om(X). 
\label{omdec}
\ee
We have introduced the bold notation ${\bf k}$ to differentiate the norm of ${\bf k}$, $k > 0$, from ${\bf k}$ itself which belongs to $(-\infty, \infty)$. 
In the $k$-representation, Eq.~\eqref{dispKGeq} gives the second order equation
\be
\left( \frac{1}{a(t)}\, \partial_t \, a(t)  \partial_t + F^2(\frac{k^2}{a(t)^2}) \right) \phi_{\bf{k}}(t) =0 \, .
\label{kmodeq}
\ee
As in all cosmological spaces, see e.g. Ref.~\cite{Macher:2008yq}, the general solution thus lives in a two-dimensional space and takes the form
\begin{equation}
\label{solphikclass}
\phi_{\bf{k}}(t) = A_{\bf{k}} \, \phi_k(t) + ( B_{\bf{-k}} \, \phi_k(t))^*. 
\end{equation}
In de Sitter, and in de Sitter only, the $k$ and $t$ dependences in Eq.~\eqref{kmodeq} can be combined in a single variable which, moreover, turns out to be the preferred momentum $P= H k \ep{-H t}$. Indeed Eq.~\eqref{kmodeq} can be rewritten as  
\be
\label{Pmodeeq}
\left(H^2 P^2\partial_P^2  + F^2(P^2)  \right) \chi(P) =0 .
\ee
This possibility is due to the presence of the \enquote{spectator} Killing field $K_t$. Whereas $K_z$ guarantees that $\bf k$-modes separate, $K_t$ tells us that Eq.~\eqref{kmodeq} is invariant under 
\begin{equation}
\label{repar}
k \rightarrow k\,  \ep{H T}, \quad
t\rightarrow t+T,  \quad
P\rightarrow P.
\end{equation}
This implies that $\phi_k(t)$ only depends on $t$ only through $P$. The same is true for the action $S_k(t) = S(t,z) - kz$, where $S(t,z)$ is solution of \eq{dispHJeq}. 

On the other hand, in the $\om$-representation, the spatial modes obey the higher order equation
\be
\left[ -(\omega+i \partial_{X} v)(\omega + i v \partial_{X})+ F^2(-\partial_X^2 ) \right] \phi_\omega(X)  = 0. 
\label{ommodeeq}
\ee
Unlike what is found in the $k$-representation, at fixed $\om$, the dimensionality of the space of solutions now depends on the dispersion relation: it is $2n$ when the highest power of $P$ in $f(P^2)$ is $2n$. In spite of this it is possible, and very instructive, to relate the solutions of  Eq.~\eqref{ommodeeq} to those of Eq.~\eqref{kmodeq}. To this end, it is useful to consider the Fourier transform,
\be
\tilde \phi_\omega({\bf{P}})=\int_{-\infty}^\infty \frac{d{{X}}}{\sqrt{2\pi}} \ep{-i {\bf{P}}{{X}}} \phi_\omega ({{X}}),
\ee
where ${\bf P}$ designates the wave vector, and $P > 0$ its norm. In the $\om,P$-representation, Eq.~(\ref{ommodeeq}) becomes
\be 
\label{ommodeeqinP}
\left[-(\omega-i H {{P}} \partial_{{P}})(\omega - i H \partial_{{P}} {{P}}) +F^2(P^2)\right] \tilde \phi_\omega  = 0 .
\ee
As in the $k$-representation, this is a second order equation (in $\partial_P$). Moreover one verifies, that $\tilde \phi_\omega$ exactly factorizes as~\cite{Brout:1995wp}
\be
\tilde \phi_\omega({\bf{P}}) = P^{-i \frac{\omega}{H}-1}\times  \tilde \chi({\bf{P}})\, , 
\label{facto}
\ee
where $ \tilde \chi$ is independent of $\om$. In addition, one also verifies that $ \tilde \chi$ obeys Eq.~\eqref{Pmodeeq}. These unusual properties are due to the other Killing field $K_z$. In fact because of Eq.~\eqref{repar}, in $(t,P)$ representation, irrespectively of Lorentz violating term $f(P^2)$, the modes trivially depend on $t$ through a delta-function: $\delta(P-H k\ep{H t})$. When working in the $(\om,P)$ representation, this implies both the factorization of Eq.~(\ref{facto}), and the fact that $\tilde \chi$ obeys Eq.~\eqref{Pmodeeq}. The extra factor of $1/P$ in Eq.~\eqref{facto} is due the Jacobian $dt/dP = -1/HP$. 

Since Eq.~(\ref{ommodeeqinP}) is second order and singular at $P=0$, the dimensionality of the space of solutions of $\tilde \phi_\om({\bf P})$ is $4$, because $\bf P$ has both signs. The physical meaning of this four-dimensional space, and its relation with the two-dimensional one found in the $k$-representation, are given below.

\subsection{Scalar product and BD vacuum} 

To complete the comparison between the solutions of Eqs.~\eqref{kmodeq} and~\eqref{ommodeeq}, we consider the conserved scalar product. It is given by~\cite{Unruh:1994je}
\be
(\Phi_1,\Phi_2) =  i \int dl \left( \Pi_1^* \Phi_2  -  \Phi_1^* \Pi_2\right) , 
\label{scalpr}
\ee
where $\Pi = - u^\mu \partial_\mu \Phi$ is the momentum conjugated to $\Phi$. The integral must be evaluated along $u_\mu dx^\mu = dt = 0$, and the line element is $dl = dX = a(t) dz$.

In the $k$-representation, for $ \Phi_{\bf{k}}= \ep{i{\bf{k}}z} \phi_{\bf{k}} $ and $\Phi_{\bf{k'}} = \ep{i{\bf{k'}}z}  \phi_{\bf{k'}}$, one has 
\be
( \Phi_{\bf{k}},  \Phi_{\bf{k'}}) = 2\pi \delta({\bf{k}}-{\bf{k'}}) \times \left[ a(t) \left( \phi^*_{\bf{k}} i\partial_t \phi_{\bf{k'}}-\phi_{\bf{k'}} i\partial_t \phi^*_{\bf{k}}\right) \right]  .
\label{scalprink}
\ee
The standard normalization $( \Phi_{\bf{k}},  \Phi_{\bf{k'}}) =2 \pi \delta({\bf{k}}-{\bf{k'}})$ imposes to work with modes $\phi_{\bf{k}}$ that have a unit positive current with respect to $a(t) i\partial_t$. When considering $\chi$ of Eq.~\eqref{Pmodeeq}, it is convenient to reexpress this condition as
\be
 H\chi i\partial_P \chi^* -H\chi^* i\partial_P \chi =1 . 
\label{unitW}
\ee
That is, the $\chi$ mode is imposed to be of unit positive Wronskian. However, because Eq.~\eqref{Pmodeeq} is second order, $\chi$ is not completely fixed by \eq{unitW}. To identify the $in$ mode which describes particles at early time, one has to impose that it behaves as the positive frequency \WKB mode at early time~\cite{birrell1984quantum}. Using Eq.~\eqref{repar} to reexpress this condition in terms of $P$, the $in$ mode $\chi_{\rm BD}$ must obey
\begin{equation}
\label{chiWKB}
\chi_{\rm BD}\underset{P\to \infty}{\sim} \frac{\ep{i \int^P {F\left(P^{\prime \, 2}\right)} \frac{dP'}{P' H}}}{\sqrt{2 \frac{F(P^2)}{ P}}}.
\end{equation}
Then the modes $\phi_{\bf k}$ with unit positive norm can all be written as 
\be
\label{kp}
\phi_{\bf{k}}(t) = \frac{1}{\sqrt{k}}\left[A_{\bf{k}} \,  \chi_{\rm BD}(P) + (B_{-\bf{k}} \, \chi_{\rm BD}(P))^*\right],
\ee
where $A_{\bf{k}}$ and $B_{-\bf{k}}$ satisfy $\abs{A_{\bf{k}}}^2-\abs{B_{-\bf{k}}}^2=1 $, and where the extra factor of $\sqrt{1/k}$ ensures that $( \Phi_{\bf{k}},  \Phi_{\bf{k'}}) =2 \pi \delta({\bf{k}}-{\bf{k'}})$ is found when $\chi_{\rm BD}$ obeys Eq.~\eqref{unitW}. The state which is vacuum with respect to $\chi_{\rm BD}$ for all values of $\bf k$, i.e., $B_{\bf k} = 0$ for all $\bf k$, is the Bunch-Davies (BD) vacuum~\cite{Schomblond:1976xc,Bunch:1977sq}.

To handle the mode identification in the $\omega$-representation, it is appropriate to work with the Fourier mode of Eq.~\eqref{facto} and to separate solutions with positive and negative values of $\bf P$. In the WKB approximation, positive norm solutions describe right moving ($U$) particles for ${\bf P} > 0$, left moving ($V$) particles for ${\bf P} < 0$, and vice versa for negative norm solutions. However, the exact solutions of Eq.~\eqref{ommodeeqinP} mix  $U$ and $V$ modes. The general solution should thus be decomposed as  
\begin{equation}
\label{omegaP}
\begin{split}
\tilde \phi_\omega({\bf{P}}) = &\left(\frac{P}{H}\right )^{-i \frac{\omega}{H}-1} \left\{ \frac{\theta({\bf{P}})}{H } \left[A_\omega^U \,  \chi_{\rm BD} + \left (B^{V}_{-\omega} \, \chi_{\rm BD} \right )^*\right] \right .\\
&+\left .  \frac{\theta(-{\bf{P}})}{H } \left[ A^V_\omega \, \chi_{\rm BD} + \left ( B^{U}_{-\omega} \, \chi_{\rm BD} \right )^*\right] \right\},
\end{split}
\end{equation}
where the $4$ coefficients weigh the initial (BD) contributions with positive (negative) norm $A$ ($B$), and with $U$ or $V$ content. In fact, the scalar product of two such modes $\Phi_\om ={\ep{- i\omega t}}  \phi_\omega, \, \Phi_{\omega'} = {\ep{- i\omega' t}} \phi_{\omega'}$ is 
\begin{equation}
\begin{split}
(\Phi_{\omega} , \Phi_{\omega'} )&=2 \pi H\,  \delta(\omega-\omega') \\
&\left( \abs{A^U_\omega}^2-\abs{B^U_{-\omega}}^2+\abs{A^V_\omega}^2- \abs{B^{V}_{-\omega}}^2\right) .
\end{split}
\end{equation}
This is exact and can be verified be expressing Eq.~\eqref{scalpr} in the $P$-representation~\cite{Balbinot:2006ua}. Imposing the positive norm condition $(\Phi_{\omega} ,  \Phi_{\omega'} ) = 2 \pi H \delta(\omega-\omega')$ on the mode basis constraints the above parenthesis to be unity. Hence, in de Sitter, irrespectively of the dispersion relation of Eq.~\eqref{disprel}, the complete set of positive norm stationary modes contains $2$ modes $\phi_\om^U,\phi_\om^V$ for $\om\,  \in  (-\infty, \infty)$. One verifies that $2n-4$ solutions of Eq.~\eqref{ommodeeq} are not asymptotically bounded in $X$, and cannot be normalized. These modes should not be used when decomposing the canonical field obeying Eq.~\eqref{dispKGeq}~\cite{Macher:2009tw}. It is interesting to notice that the completeness of the stationary modes follows from the completeness of the Mellin transform~\cite{morse1953methods}, in a manner similar that the completeness of the homogeneous mode basis follows from that of the Fourier transform, see the end of Appendix~\ref{group} for more details. With this remark,  we have verified that the set of the asymptotically bounded solutions of Eq.~\eqref{ommodeeq} matches that of the solutions of Eq.~\eqref{kmodeq}. 

To conclude this section, we point out that the $S$-matrix in the $k$-representation factorizes into 2-mode sectors containing particles with opposite wave vectors $\bf k$, because the space in homogeneous. Instead, in the $\om$-representation, the  $S$-matrix  factorizes in different sectors with $\om > 0$, each of them being a 4-mode sector which contains two $U$ modes $\phi_\om^U, (\phi_{-\om}^U)^*$, and two $V$ modes $\phi_\om^V, (\phi_{-\om}^V)^*$. The $4 \times 4$ character of the  $S$-matrix in this representation results from the composition of the cosmological mixing of $U$ and $V$ modes with the stationary mixing of modes of opposite frequency, see Sec.~\ref{Smatrix} for details.

\subsection{The two Hamiltonians} 
\label{Hamiltonians}

In preparation for the analysis of the stability of the BD vacuum, we study the Hamiltonian of our dispersive field. We first point out that the fields $u$ and $K_t$ define two different Hamiltonian functions, that we call, respectively, $H_u$ and $H_t$. Using the conjugated momentum $\Pi = - u^\mu \partial_\mu \phi $ and the Lagrangian density $L = - \Phi \hat O \Phi /2$, where $\hat O \Phi = 0$ is Eq.~\eqref{dispKGeq}, they are respectively given by 
\be
\label{prefE1}
\begin{split}
H_u &\doteq \int dl (  \Pi \, \partial_t|_z \phi  - L), \\
H_t &\doteq \int dl ( \Pi \,  \partial_t |_X \phi - L),
\end{split}
\ee
where we recall that $\partial_t |_z= -u^\mu  \partial_\mu  $ and $\partial_t |_X = -K_t^\mu  \partial_\mu $. $H_u$ thus engenders time translations at fixed $z$, while $H_t$ does it at fixed $X$. In de Sitter space, $H_u$ and $H_t$ differ because $u = K_t - v s \neq K_t$ since the flow $v= HX$ does not vanish. In Minkowski space endowed with a Cartesian $u$ field, which is obtained in the limiting case $H \to 0$, the two Hamiltonians coincide since $u \to K_t$ when  $H \to 0$. This implies that in de Sitter $H_t $ and $H_u$ share the properties that the Hamiltonian possessed in Minkowski space. 

On the one hand, the stationary $ H_t$ 
\be
\begin{split}
\label{HtX}
 H_t= \frac12 \int_{-\infty}^\infty dX &\left\{ (\partial_t\phi)^2 + m^2\phi^2 + (1 - v^2) (\partial_X \phi)^2\right . \\
 &+\left .  \phi f(- \partial_X^2) \phi \right\} ,
\end{split}
\ee
is conserved for all dispersion relations. However, for both Lorentz invariant theories and dispersive ones with $f(P^2)\geq 0$, it is {\it not} positive definite precisely because $K_t$ is space like outside the horizons. (Recall that its norm is $- K_t^2  =  1  - v^2$.) Notice also that when working in the $P$ representation one easily verifies that the last term in $H_t$ is positive definite for $f> 0$. For dispersive theories with $f < 0$, such as phonons in Helium$^4$, the density of $H_t$ becomes negative where $v$ reaches the critical Landau velocity~\cite{Jacobson:1999zk,Pitaevskii1984}. In any case, in de Sitter,  when using the stationary modes of \eq{omegaP}, $H_t$ can be decomposed as $\int_{0}^\infty d\om H_\om $, where
\be
\label{omE}
\begin{split}
 H_\om &= {\omega} \left(\abs{A_\om^U}^2 + \abs{A_\om^V}^2 - \abs{B_{-\om}^U}^2 - \abs{B_{-\om}^V}^2 \right)\, .
\end{split}
\ee
$H_t$ is thus manifestly conserved and not positive definite.

On the other hand, the cosmological $H_u$ can be decomposed as $H_u =\int_{-\infty} ^\infty d{\bf k} H_{\bf k}$ where
\be
\label{prefE}
\begin{split}
 H_{\bf k}(t) &=  \frac{a(t)}{ 2} \left( \abs{\partial_t \phi_{\bf k}(t)}^2 + F^2(\frac{k^2}{ a(t)^2}) {\abs{\phi_{\bf k} (t)}}^2 \right) .
\end{split}
\ee
$H_u$ is thus positive definite whenever $F^2 > 0$. (Notice that theories with $F^2 < 0$ are dynamically unstable even in Minkowski space.) However, $H_t$ is not conserved because $ d \ln a /dt = H \neq 0$. The nonconservation of $H_{\bf k}(t)$ engenders nonadiabatic transitions~\cite{Massar:1997en} which describe pair creation of quanta of opposite $\bf k$, see Sec.~\ref{betacosmo} for a particular example. When using Eq.~\eqref{kp}, the time dependence of $H_k$ can be entirely expressed through $P = Hk e^{-Ht}$ as
\be
\begin{split}
 H_{\bf k}(t)\! &=\! \frac{ \abs{ A_{\bf k}}^2+ \abs{ B_{ -\bf k}}^2 }{2 P }\! \left({\abs{HP\partial_P \chi_{\rm BD}}}^2 + F^2(P^2)\,  {\abs{\chi_{\rm BD}}}^2 \right) \\
 &+ {\rm Re}\left\{  \frac{A_{\bf k} B_{ -\bf k}}{P } \, \left( (HP\partial_P \chi_{\rm BD})^2 + F^2(P^2)\,  \chi_{\rm BD}^2 \right)   \right\}. 
\end{split}
\ee
We also see that when imposing $\abs{A_{\bf{k}}}^2-\abs{B_{-\bf{k}}}^2=1 $, the minimization of $H_k$ (more precisely, its integral over one period) implies $B_{-\bf{k}} = 0$. This is the classical equivalent of saying that the BD vacuum is the lowest energy state with respect to the preferred frame field $u$.

To conclude, we clearly see the complementary roles played by the Hubble constant $H$. In the stationary representation, it is responsible for an {\it energetic instability}, i.e., for a conserved Hamiltonian $H_t$ unbounded from below. Instead in the homogeneous representation, $H$ is responsible for the time dependence of the positive definite $H_u$, which engenders pair creation, i.e., a {\it vacuum instability}. These two properties are valid for all dispersion relations, and therefore they also apply to Lorentz invariant theories. This is a reminder that field theories in de Sitter space, and in black hole metrics, are threatened by dynamical instabilities, i.e., complex frequency modes~\cite{Damour:1976kh,Coutant:2009cu}. In addition, as argued below, violations of thermodynamical laws are also related to an energetic instability. 

\section{The consequences of Lorentz violations} 
\label{thermal}

In de Sitter space, when considering Lorentz invariant fields, the Bunch-Davies vacuum possesses many remarkable properties. On one hand, it is homogeneous and stationary, and on the other hand, it is an Hadamard state. In fact, it is the only stationary Hadamard state~\cite{Schomblond:1976xc,Joung:2006gj}. In addition, it can be shown that all other Hadamard states flow towards the BD vacuum. By this we mean that the $n$-point functions evaluated in these states flow towards the corresponding one evaluated in the BD vacuum. In this sense, the BD vacuum is the only stable regular state. Finally, when evaluated in the static patch $\vert HX\vert < 1$, the $n$-point functions are all thermal~\cite{Kay:1988mu}. They indeed obey the double KMS condition: they are periodic in imaginary time with period ${2\pi}/{H}$, and they are analytic in the strip $0<{\rm Im}\, t < {2\pi}/{H}$. 

When considering  dispersive fields,  we shall see that, for free fields at least, the BD vacuum still satisfies all these properties, save the very last. In fact, even though the periodicity in  Im $t$ is still exactly found, the analyticity in the strip is always lost when there is high frequency dispersion. This means that the BD vacuum is no longer a thermal state. 

\subsection{Stationarity and periodicity}

Since we work with free fields and since the BD vacuum is a Gaussian state, we only need to consider the $2$-point function. When using the settings of the former section, the Wightman function in the BD vacuum can be written as~\cite{birrell1984quantum}
\begin{equation}
\label{greenfunction}
\begin{split}
&G_{\rm BD}({{z}},t,{{z}}_0,t_0)  =\\
&\int_{-\infty} ^{\infty} d{\bf{k}} \frac{\ep{i{\bf{k}}(z-z_0)}}{2 \pi k} \chi_{\rm BD}\!\left(k H\ep{-Ht}\right) \, \chi^{*}_{\rm BD}\!\left(k H \ep{-Ht_0}\right) .
\end{split}
\end{equation}
When considered at fixed $t_0$ and $t$, this function is manifestly homogeneous. It is also stationary, when considered at fixed $X_0 = a(t_0)\, z_0 $ and $X = a(t)\, z $. Indeed in terms of $P = k/a(t)$, one gets 
\be
\begin{split}
G_{\rm BD}({{X}},t,{{X}}_0,t_0) &= \int_0^\infty dP  \frac{i \sin(PX - PX_0 e^{H(t-t_0)})}{  \pi P}\\
&\chi_{\rm BD}(P)\, \chi_{\rm BD}^{*}(P e^{H(t-t_0)}) ,
\label{gfP}
\end{split}
\ee
which is a function of $t - t_0$ only. Hence, for all dispersion relations imposed in the cosmological frame, the BD vacuum is both stationary and homogeneous, as it is for relativistic fields. More surprisingly, $G_{\rm BD}$ is also periodic in the imaginary time lapse, with the usual period $2\pi/H$, exactly as for Lorentz invariant fields, and for thermal functions. 

We expect that homogeneity, stationarity and the above periodicity will be exactly preserved when considering the $n$-point functions of interacting fields evaluated in the BD vacuum, because these properties are protected by the affine group of Eq.~\eqref{alg}. In other words, the $n$-points functions will always be invariant under this subgroup. 

\subsection{Thermality} 
\label{thermbroke}

For Lorentz invariant theories, it has been shown by Gibbons and Hawking~\cite{Gibbons:1977mu} that freely falling observers immersed in the BD vacuum detect a thermal bath with a temperature $T_H =\frac{H}{2\pi }$ in natural units $k_B = \hbar = c = 1$. It was also understood that, when restricted to the static patch $-1 <  HX < 1$, the reduced density matrix of any quantum field theory (interacting or not) is a thermal state at that temperature. Interestingly, this result is always violated for dispersive fields. 

To demonstrate this, we consider particle detectors which follow orbits that are stationary with respect to the Killing field $K_t$. Because $K_t^2 = - 1 + H^2 X^2$, the only stationary orbits are at fixed $X$, and with $\vert HX\vert < 1$ when they are timelike. The detector transition rate of spontaneous excitation $R_-$ (de-excitation $R_+$) is proportional to~\cite{Unruh:1976db,Brout:1995rd} 
\begin{equation}
\begin{split}
R_{\pm} (\omega,{{X}})= & \int_{-\infty}^{\infty} H dt  \, \ep{\pm i \omega t} \, G_{\rm BD}({{X}},t; {{X}}, 0).
\label{Rs}
\end{split}
\end{equation}
To relate the detector energy gap $\Delta E > 0$ to the Killing frequency $\om$, one must take into account the $X$ dependent redshift factor ($\Delta E = \om / \sqrt{1-H^2 X^2}$) coming from the detector's kinematics, which also enters in Tolman law $T_{\rm loc} (X)= T_{\rm gl} / \sqrt{1-H^2 X^2}$ relating the local temperature to the globally defined one~\cite{Misner1973,Tolman:1930zza}. In what follows, we shall work with the globally defined temperature and with $\om$. To study the deviations from thermality, it is convenient to use the temperature function $T_{\rm gl}(\om,X)$ defined by
\begin{equation}
\begin{split}
\label{tempdef}
\frac{R_-(\omega,X)}{R_+(\omega,X)}&= \ep{ - {\omega}/{T_{\rm gl}(\omega,X)}}.
\end{split}
\end{equation}
For relativistic fields, one has $T_{\rm gl}(\om,X)= T_H = H/2\pi$, for all $\vert HX\vert < 1$ and for $0< \om < \infty$, in accord with Tolman law and the Planck spectrum. To compute $T_{\rm gl}$ in the presence of dispersion, we shall use the fact that the rates are given by 
\begin{equation}
\begin{split}
R_\pm(\omega,{{X}}) = \abs{\phi^{{\rm BD}, U}_{\pm\om}({{X}})}^2+\abs{\phi^{{\rm BD}, U}_{\pm\omega}(-{X})}^2, 
\end{split}
\label{Rs2}
\end{equation}
where $\phi^{{\rm BD}, U}_{\om}$ is the positive norm BD mode that is initially right moving, i.e., $A^U = 1, \,  A^V = B^U = B^V = 0$ in Eq.~\eqref{omegaP}. Similarly, $(\phi^{{\rm BD}, U}_{-\om})^*$ is given by the negative norm, negative frequency, mode: $B^U= 1$. In the above equation we have used the symmetry $X \to - X$ to express the contribution of the left moving $V$-mode evaluated at $X$ as that of the right $U$-moving one at $-X$. Explicitly, $\phi^{{\rm BD}, U}_{\om}$ is 
\begin{equation}
\label{amp0}
\begin{split}
\phi^{{\rm BD}, U}_{\om}({{X}})= \int_0^\infty \frac{dP}{\sqrt{2 \pi} H} \ep{i P X} \left (\frac{P}{H}\right )^{- i\omega/H -1}  \chi_{\rm{BD}}(P) ,
\end{split}
\end{equation}
where $\chi_{\rm{BD}}$ obeys \eq{Pmodeeq}. To prove that thermality is violated it is sufficient to work with $X= 0$ and to consider very high frequencies $\omega/\Lambda \gg 1$. In this limit the integral is dominated by high values of $P$, and therefore by the leading term of the dispersion relation, that we parametrize here by 
\be
f_n(P^2)=\frac{P^{2n}}{\Lambda^{2n-2}}. 
\label{fn}
\ee
In the high $P$ regime, the \WKB expression of Eq.~\eqref{chiWKB} offers a reliable approximation of $ \chi_{\rm{BD}}$. Hence, up to irrelevant constants, one gets
\be
\phi^{{\rm BD}, U}_{\om}({{X}}= 0)  \approx \int \frac{dP}{P} P^{- i\frac{\omega}{H} - \frac{n-1}{2}} \, e^{i P^n} . 
\ee
Using $Q= P^n$ as integration variable, one obtains a $\Gamma$ function, namely
\begin{equation}
\begin{split}
\phi^{{\rm BD}, U}_{\om}({{X}}= 0)  &\approx{{e}^{{{ \omega\,\pi }/{2nH} }}} \times \Gamma  \left( {-\frac {i\omega}{n H}} - \frac{n-1}{2n} \right) .
\end{split}
\end{equation}
Using this result in Eq.~\eqref{Rs2}, Eq.~\eqref{tempdef} gives
\begin{equation}
\begin{split}
T_{\rm gl}(\om \gg\Lambda, X = 0) = n \, \frac{H}{2\pi},
\label{nT}
\end{split}
\end{equation}
i.e., $n$ times the standard temperature $T_H$. 

Instead, for $\omega/\Lambda \ll 1$ and $H/\Lambda \ll 1$, $T_{\rm gl}(\om, X = 0)$ reduces approximatively to the standard temperature $T_H$~\cite{Coutant:2011in}. Hence the BD vacuum is no longer thermal. This result is nontrivial since $G_{\rm BD}$ of Eq.~\eqref{gfP} is still periodic in imaginary time, with the standard period. From the above equations and from $P \sim e^{-Ht}$, one understands that the power $n$ of Eq.~\eqref{fn} reduces the domain of analyticity of $G_{\rm BD}$ in Im $  (t)$ by a factor of $n$. Indeed since $\chi_{\rm BD} \sim  \ep{i P^n }$ for large $P$, the integral in Eq.~\eqref{gfP} contributes as ${1}/({1-\ep{n H  t}})$ which is analytic in the reduced strip $0< {\rm {Im}} (t)< 2\pi /n H$ only. The observation that $T_{\rm gl}(\om, X = 0) = n \, T_H$ for $\om/\Lambda \gg 1$ shall be verified, for $n=2$, in a exactly solvable model in Sec.~\ref{thermidev}. 

\subsection{Regular states and stability}
\label{Hadamard}

In this section, we show that some of the ingredients of event horizon thermodynamics~\cite{Gibbons:1977mu,Jacobson:2003wv} are still present when adding high frequency dispersion. Namely, we show, firstly that the BD vacuum is the only stationary state which is regular and, second, that the other regular states flow towards the BD state. (As explained below, the notion of regular states should be understood as the generalization of Hadamard states in the presence of short distance dispersion.) Hence for free fields at least, the BD state is the only stable state. To prove these claims we shall use concepts that are common to Lorentz invariant and dispersive fields. 

Before proceeding, let us discuss our criterion of stability. We say that the BD vacuum is stable because, at large time, observables computed in nearby states converge towards those evaluated in the BD vacuum. Hence, for these observables, the perturbed states will be asymptotically indistinguishable from the BD vacuum. This flow is often referred to as a cosmic no hair theorem~\cite{Marolf:2010nz,Marolf:2011sh,Hollands:2010pr} as it closely follows the Price's no hair theorem~\cite{Misner1973}. We adopted this criterion because there is no stationary Killing field which is globally timelike in de Sitter. As a consequence, there is an {\it energetic} instability, see Eq.~\eqref{omE}, which means that stability cannot be deduced from a spectrum bounded from below. It is worth mentioning that to study the thermalization in interacting quantum field theories, the flow towards stationary thermal states is established in Ref.~\cite{Giraud:2009tn} by studying some $n$-point functions. Even though the purity of the initial state is preserved by the Hamiltonian evolution, after a while, these functions become indistinguishable from thermal ones. In that case as well, the stability of the state is thus inferred from the flow of some observables, rather than from the evolution of the state itself. 

Since high frequency dispersion modifies the short distance behavior of the $2$-point function, we first need to define what we mean by \enquote{regular states} because the standard definition of Hadamard state is precisely based on this behavior~\cite{birrell1984quantum}. In homogeneous cosmological spaces, this difficulty can be overcome because one can rephrase the standard definition in terms of an adiabatic expansion of the solutions of Eq.~\eqref{kmodeq} at fixed $\bf k$. Since these new terms are common to both Lorentz invariant and dispersive field, one can implement the subtraction procedure to dispersive fields. Let us recall the key elements, for more details, see Ref.~\cite{LopezNacir:2005db}. In de Sitter, because of Eq.~\eqref{alg}, the adiabatic expansion can be done in terms of a single mode $\chi$, solution of Eq.~\eqref{Pmodeeq}, and of unit Wronskian, see Eq.~\eqref{unitW}. This expansion generalizes Eq.~\eqref{chiWKB} and is best expressed as 
\begin{equation}
 \chi_{\rm adiab}= \sqrt{\frac{1}{2 W(P)}} \ep{i\frac{1}{H} \int^{P} W(P') dP'}, 
\ee
where $W$ obeys the nonlinear equation
\be
W^2 = \frac{F^2(P^2)}{P^2}-\frac{H^2}{2} \left( \frac{\partial_P^2 W}{W}-\frac{3}{2} \frac{(\partial_P W)^2}{W^2} \right),
\label{WF}
\ee
and where $F^2$ determines the dispersion relation in Eq.~\eqref{disprel}.

When working with Lorentz invariant fields in $D$ dimensions, the first $1+ D/2$ terms in a iterative solution of Eq.~\eqref{WF} should be taken into account when determining the $1+ D/2$ quantities that need to be subtracted. This guarantees that the renormalized stress tensor evaluated in the BD vacuum is finite in cosmological spaces, and thus in de Sitter. This is not a surprise since $\chi_{\rm BD}$ and $\chi_{\rm adiab}$ obey the same condition for $P\to \infty$. Hence their differences develop at finite $P$, and because of the expansion $H$. When working with dispersive fields, this finiteness is still found  when $F^2$ is positive, sufficiently regular, and grows faster that $P^2$ for $P \to \infty$. Indeed the higher the power $n$ in the leading term of \eq{WF}, the more suppressed are the next order terms in the adiabatic expansion~\cite{LopezNacir:2005db}. For instance, in two dimensions, for $F^2_n \sim P^{2n}/\Lambda^{2n -2}$, with $n\geq 2$, the second quantity which is usually subtracted in the stress tensor is already finite. It can thus be either subtracted or not.\footnote{
As a result, in Ref.~\cite{LopezNacir:2005db}, it is proposed to subtract only the first term. We claim instead that the first two terms should be subtracted, as done when dealing with Lorentz invariant fields. Indeed only this choice guarantees that the stress tensor would remain finite when taking the limit $\Lambda\to \infty$. In addition, in our proposal, the two manners to consider $\Lambda$ become compatible. $\Lambda$ can be either seen as a (Lorentz violating) regulator to be sent at $\infty$ when computing observables, or as a physical finite ultraviolet parameter, but which enters suppressed in observables.}
In either case, in the BD vacuum of de Sitter, the renormalized values of $\rho = u^\mu u^\nu T_{\mu \nu}$ and $\Pi= s^\mu s^\nu T_{\mu \nu}$ are constant in space and time, while the flow $J = u^\mu s^\nu T_{\mu \nu}$ vanishes. 

We now consider the change of the stress tensor with respect to that of the BD vacuum when working with some (possibly mixed) state $\Psi$ described by  the density matrix $\hat \rho_\Psi$. For free fields, this change is determined by the difference of the $2$-point functions $\delta G_\Psi= G_\Psi - G_{\rm BD}$. This difference can be expressed in terms of the positive norm BD modes $\phi_{\bf k}^{\rm BD}$ as 
\begin{equation}
\label{green}
\begin{split}
\delta G_\Psi&= 2 {\rm Re}\int \frac{{d{\bf{k}} d{\bf{k'}}}}{2\pi} \left[  \ep{i ({\bf{k}} {{z}}-{\bf{k'}} {{z'}})} \phi_{\bf k}^{\rm BD}(t) \right .\\
&\left .\left\lbrace n_\Psi({{\bf{k}},{\bf{k'}}})\,   (\phi_{\bf k'}^{\rm BD}(t'))^* +c_\Psi({{\bf{k}},{\bf{k'}}})\,   \phi_{\bf k'}^{\rm BD}(t')\right\rbrace  \right] ,  
\end{split}
\end{equation}
where $n_\Psi({{\bf{k}},{\bf{k'}}})$ and $c_\Psi({{\bf{k}},{\bf{k'}}})$ are expectation values of normal ordered products of BD destruction and creation operators $a_{\bf{k}}, a^{\dagger}_{\bf{k'}}$: 
\begin{equation}
\begin{split}
n_\Psi({{\bf{k}},{\bf{k'}}} )= {\rm Tr} \left[\hat \rho_\Psi \, a^{\dagger}_{\bf{k'}} a_{\bf{k}} \right],
\quad
c_\Psi({{\bf{k}},{\bf{k'}}}) = {\rm Tr} \left[\hat \rho_\Psi \,  a_{\bf{k}} a_{-{\bf{k'}}}\right]. 
\end{split}
\end{equation}
They respectively encode the power spectrum and the coherence of $\hat \rho$ at the Gaussian level~\cite{Campo:2005sy}.

To establish the stability of the BD vacuum, we first point out that the other stationary states are all singular. The reason comes from the fact that the stationarity of $G_\Psi$ implies that, irrespectively of $c_\Psi$, ${\bf{k}}\times n_\Psi({{\bf{k}},{\bf{k'}}} )$ only depends on the ratio ${{\bf{k}}/{\bf{k'}}}$. Therefore, the change of the expectation value of $H_u$ of Eq.~\eqref{prefE1}  with respect to the BD vacuum, necessarily diverges because the contribution of high ${\bf k}$ is not suppressed enough, as $ n_\Psi({{\bf{k}},{\bf{k}}} )\propto 1/k$. This is true for dispersion functions $F(P) \geq \epsilon >0 $ for $P\to \infty$, and therefore true for Lorentz invariant theories. This generalizes the fact~\cite{Joung:2006gj} that the $\alpha$-vacua, which are invariant under the full de Sitter group and therefore stationary, are all singular, save the BD vacuum. 

In our second step we consider states that describe, at some initial time, {\it local} perturbations containing a {\it finite} number of BD particles: $N_{\rm tot} = \int d{\bf k}\, n_\Psi({{\bf k},{\bf k}}) < \infty$. Moreover, to be able to handle all dispersion relations at once, we suppose that there exists a cut off wave number $k_{\max}$ above which the number of particles decreases exponentially, i.e.
\begin{equation}
\label{nottoofar}
n_\Psi({{\bf k},{\bf k}}) \leq \ep{- b\,  k },  \quad \forall k>k_{\max},
\end{equation}
with $b > 0$. Then Schwartz inequalities and the hermiticity of $\hat \rho_\Psi$ implies the following inequalities generalizing those of Ref.~\cite{Campo:2005sy} 
\begin{equation}
\label{campo} 
\abs{n_\Psi({{\bf k'},{\bf k}})}^2 \leq \abs{n_\Psi({{\bf k},{\bf k}})}
\abs{n_\Psi({{\bf k'},{\bf k'}})},
\ee 
and 
\be
\begin{split}
 &\abs{\int d{\bf k_1} f_{{\bf k_1}} c_\Psi({{\bf k_1},{\bf k}})}^2 
\leq {n_\Psi({{\bf k},{\bf k}})} \\
&\hspace{1cm}\left [\int d{\bf k_1} d{\bf k_2} 
f_{{\bf k_1}} f_{{\bf k_2}}^* (n_\Psi({{\bf k_1},{\bf k_2}})+\delta({\bf k_1}-{\bf k_2}))\right ],
\end{split}
\end{equation}
for all test functions $f_{\bf k} \in \mathbb{C}$.

Using these inequalities one can study the behavior of Eq.~\eqref{green} at large time. Since there is a momentum cutoff $k_{\max}$, at large time only low momenta $P$ matter. Hence for all dispersion relations of Eq.~\eqref{disprel}, the dominant term is the mass term. One should then distinguish massive fields with $m > H/2$ [see Eq.~\eqref{mum} for the origin of this condition] from massless fields. At this point, one also needs to consider the pair creation amplitudes relating the initial BD mode $\phi^{\rm BD}_{\bf k}$ [obeying Eq.~\eqref{chiWKB}] to the out mode $\phi^{out}_{\bf k}$ defined at low momentum. Using the techniques of Ref.~\cite{Massar:1997en} and the fact that Eq.~\eqref{Pmodeeq} is second order for all $F^2$, we can verify that for both $m > H/2$ and $m = 0$, the $\alpha_k, \beta_k$ coefficients of Eq.~\eqref{Bogok} are bounded for dispersion relations with $F^2 > 0$. Using this result, at large time and for massive fields, one finds that 
\be
\delta_\Psi G < \ep{-H (t+t')/2},
\label{expondec}
\ee
 i.e., $\delta_\Psi G$  decreases exponentially in time. This implies that the changes of density, current and pressure with respect to the BD vacuum $\delta \rho_\Psi = u^\mu u^\nu \delta T_{\mu \nu}^\Psi$, $\delta J_\Psi = u^\mu s^\nu \delta T_{\mu \nu}^\Psi$ and $\delta \Pi_\Psi = s^\mu s^\nu \delta T_{\mu \nu}^\Psi$ also flow exponentially to $0$. On the other hand, when $m=0$, at large times the dispersive modes become conformally invariant, i.e., proportional to $e^{i {\bf k} z - i k \eta}$ where $\eta \propto e^{-Ht}  $ is the conformal time. As a result, both scalar derivatives $u^\mu \partial_\mu \delta G_\Psi$ and $s^\mu \partial_\mu \delta G_\Psi$ flow to $0$ as in \eq{expondec}. This implies that $\delta \rho_\Psi$, $\delta J_\Psi$ and $\delta \Pi_\Psi$ also flow exponentially fast to $0$.

In conclusion, we have shown that for all dispersion relations, the mean stress tensor Tr$(\rho_\Psi\, T_{\mu\nu})$ computed with an arbitrary localized state containing a finite number of BD quanta flows towards that computed in the BD vacuum. This follows from the cosmological expansion $a \sim e^{Ht}$ which redshifts the momenta $P \sim k e^{-Ht}$, and dilutes the particles. In our proof we have used the condition of Eq.~\eqref{nottoofar} because, for all polynomial dispersion relations, it guarantees that the change of the stress tensor with respect to its value in the BD vacuum is finite. Less restrictive conditions, and therefore larger set of states, can certainly be used once having chosen some class of dispersion relations. One could also relax the condition that the perturbation is local. However a detailed study of these extensions goes beyond the scope of this paper. 

\subsection{Discussion and lessons for black holes} 

For all dispersion relations, we showed that the BD vacuum is the only state which is stationary, regular, and stable. For Lorentz invariant fields, this stability goes together with the fact that the BD vacuum is a thermal state when restricted to the static patch, $\vert HX \vert < 1$. Therefore, the flow of nearby states towards the BD vacuum can be meaningfully considered as a no hair theorem compatible with the laws of event horizon thermodynamics applied to de Sitter. For dispersive fields instead, there is a tension because we have demonstrated that the BD vacuum is no longer in thermal equilibrium when probed by static particle detectors. 

Before considering possible consequences~\cite{Dubovsky:2006vk,Eling:2007qd} of this loss, it is important to identify its origin. It can be traced to the loss of the positivity of $\hat H^{\rm stat.}_t$, the stationary Hamiltonian operator of \eq{HtX} {\it restricted} to the static patch, i.e., for $\vert HX \vert < 1$. To explain this, let us first recall the key properties found with relativistic fields~\cite{Unruh:1976db,birrell1984quantum,Brout:1995rd}. For these fields, there exists a complete set of (positive norm) negative frequency modes $\phi_{-\om}$ in which all modes identically vanish in the static patch, see Eq.~\eq{phiout} for the massless case. As a result, only positive frequency modes live in the patch, and the spectrum of $\hat H^{\rm stat.}_t$ is bounded from below.

For dispersive fields, this property is lost because the corresponding negative frequency modes no longer exactly vanish for $\vert HX \vert < 1$. For instance, for superluminal dispersion, they are decaying in this domain, see e.g., Ref.~\cite{Coutant:2011in} for details. When $\hat H^{\rm stat.}_t$ it is no longer bounded from below, given a certain energy in the static patch, it is no longer meaningful to look for the maximum of the entropy. 
In other words, one is lacking a necessary condition for the ordinary second law of thermodynamics (OSL) to hold. In this we do not agree with the claim \enquote{Only the validity of the GSL is in question}~\cite{Eling:2007qd}, where GSL refers to the generalized second law involving black holes. The fact that the violations of thermality are neither necessarily related to black holes, nor to event horizons, is clear when considering subluminal dispersion relations. In that case, as pointed out after \eq{HtX}, when $v$ exceeds the Landau critical velocity, $H_t$ is no longer bounded from below. As a result, there exist stationary metrics without Killing horizons where there is pair creation of quanta of opposite frequency. (An interesting illustration of this can be found in Ref.~\cite{Pitaevskii1984}.) In these cases, the ordinary zeroth law will be violated, i.e., the spectra will not be thermal. Notice that the importance of the violations will depend on the intensity of the mixing of modes with opposite frequency (which governs the instability of the system). When the UV scale $\Lambda$ is much higher than the typical value of spatial gradient $\partial_X v$, this mixing could be strongly suppressed, and therefore the violations of the thermodynamical laws accordingly so (because the system will be long living). 

Having clarified these aspects, we now point out that in de Sitter the nonthermal behavior manifests itself in the $\om$-representation, whereas the stability proof heavily used time dependent effects in the  homogeneous $k$-representation. However, we know that these two descriptions ought to be compatible with each other. The lesson therefore seems to be that in the presence of ultraviolet dispersion, the BD vacuum is still an equilibrium state, albeit with unusual (nonthermal) properties when considered in the stationary picture. So instead of viewing the violations of thermality as an indication that horizon thermodynamics might no longer exist, these laws could still exist, albeit in some modified and still unknown guise.

Since the flow towards the BD vacuum of \eq{expondec} is engendered by the cosmological expansion, $a \sim e^{Ht}$, it seems a priori difficult to apply the stability proof to black hole backgrounds, because these are stationary. However, the effects of dispersion in black hole backgrounds turn out to be essentially the same as in de Sitter when the surface gravity $\kappa= H \ll \Lambda$ (see Appendix~\ref{BHdSc}). Hence, there exist good reasons to believe that what applies to de Sitter could also apply to black holes. 

Let us present here the essential aspects of this correspondence. First, when the preferred frame is freely falling, the commutator of $u$ and $s$ obeys Eq.~\eqref{commus} which is the generalization of  Eq.~\eqref{usc}. Equation~\eqref{commus} in turn implies Eq.~\eqref{ptau} which gives, near the horizon, $P \sim e^{-\kappa t}$ both for outgoing and infalling modes, as if they were propagating in an expanding de Sitter universe. Second, at the level of the quantum theory, the deviations from the relativistic black hole spectrum are governed by the quantity $D$ of Eq.~\eqref{usnhr} which controls the spatial extension of the near horizon region which can be mapped on a de Sitter space endowed with a preferred cosmological frame. This shows that the black hole-de Sitter correspondence is not only qualitative, but quantitatively determines the spectral deviations. Third, when the preferred frame is not freely falling, the above analysis possesses a generalization which is briefly described in footnote~\ref{gamma}. On this basis, two alternative physical scenarios can be envisaged when dealing with a dispersive theory of gravity, such as Einstein-aether~\cite{Eling:2004dk} or Horava gravity~\cite{Horava:2009uw}. Either black holes are dynamically unstable in this theory, and there is no question of thermodynamical laws when the life time (inverse growth rate) of the instability is sorter or comparable to time scale of the processes under study. Or they are stable, and one can conjecture that the modified black hole thermodynamical laws will be quantitatively the same as those applying to de Sitter.

\section{Quartic superluminal dispersion}
\label{pfour}

It is of value to explicitly compute the modifications of the observables which are due to high frequency dispersion. In de Sitter there are \textit{a priori} two types of observables: first, the pair creation rates which are due to cosmological expansion, and second the thermal-like response of stationary particle detectors. In Sec.~\ref{Smatrix} we shall study a third type of observables, namely asymptotic pair creation rates in the $\om$-representation, which combines the former two phenomena.

To get analytical expressions, we consider the quartic superluminal dispersion, i.e., $f = {P^4}/{\Lambda^2}$. In this case, the general solution of Eq.~\eqref{Pmodeeq} is given by 
\begin{equation}
\label{solchi}
\begin{split}
\chi &= \frac{C}{\sqrt {p}} {{\rm \bf M}\left(\frac{-i\lambda}{4} ,\frac{i \mu}{2},\,{\frac {i p^{2}}{ \lambda }}\right)} +\frac{D}{\sqrt {p}} {{\rm \bf W}\left(\frac{-i\lambda}{4} ,\frac{i \mu}{2},\,{\frac {i p^{2}}{\lambda }}\right)},
\end{split}
\end{equation}
where $C$ and $D$ are the two integration constants, and where ${\rm \bf M}$ and ${\rm \bf W}$ are two Whittaker functions, see Ch.13 in Ref.~\cite{Abramowitz}. For simplicity, we introduced the adimensional quantities $ \mu= \sqrt {\frac{m^2}{H^2} -\frac14}$, $\lambda=\Lambda/H$, $p=P/H$. Using Eq.~\eqref{chiWKB} to characterize the initial large $P$ behavior, the unit Wronskian BD mode, when complex conjugated, is given by~\cite{Macher:2008yq,LopezNacir:2005db} 
\begin{equation}
\begin{split}
\label{chist}
\chi_{\rm BD}^* = \sqrt{\frac{\lambda}{2 p}}\, \ep{\frac{-\pi \lambda}{8 }} \, {{\rm \bf W}\left(\frac{-i\lambda}{4} ,\frac{i \mu}{2},\,{\frac {i p^{2}}{\lambda }}\right)}.
\end{split}
\end{equation}

\subsection{Cosmological pair creation rates} 
\label{betacosmo}

To get the pair creation rates, we need to identify the combination of ${\rm \bf M}$ and ${\rm \bf W}$ that corresponds to the final mode $\chi_{out}$. As Eq.~\eqref{chiWKB} does not offer a reliable approximation for $P \to 0$, the identification should be done using the cosmological time $t$. Using  Eq.~\eqref{kmodeq}, one finds that asymptotic positive norm solutions are proportional to $e^{- i \mu H t}$ at large $t$. When $m < \frac{H}{2} $, $\mu$ is imaginary and the modes grow or decay at large time~\cite{Joung:2006gj}. Hence it is not possible to define asymptotic $out$ modes. When $m > \frac{H}{2} $, there is no difficulty: when reexpressing  $e^{- i \mu H t}$ in terms of $P \propto e^{-Ht}$, one gets 
\begin{equation}
\begin{split}
\chi _{out}& \underset{p\to 0}{\sim}\frac{p^{\frac12+i{\mu}}}{\sqrt{2 {\mu} }}.
\label{mum}
\end{split}
\end{equation}
Using this behavior, the positive unit Wronskian $out$ mode is found to be
\begin{equation}
\begin{split}
\chi _{out}=  {\left(-i\lambda\right)^{\frac{1+ i  \mu }{2}}  } \sqrt{\frac{1}{2 \mu p}} {{{\rm \bf M}\left(\frac{-i\lambda}{4} ,\frac{i \mu}{2},\,{\frac {i p^{2}}{\lambda}}\right)}}.
\end{split}
\end{equation}
The $in-out$ Bogoliubov transformation is given by 
\be
\chi_{out}= \alpha_{k} \chi_{\rm BD} +\beta_{k} \chi^*_{\rm BD}\, .
\label{Bogok}
\ee 
We put a subscript $k$ to the above ($k$-independent) coefficients  to remind the reader that all these calculations are done in the $k$-representation. Using Sec 13.1 in Ref~\cite{Abramowitz}, one finds, 
see Appendix~B.2 in Ref.~\cite{Macher:2008yq}, 
\begin{equation}
\label{betakmassive}
\begin{split}
\abs{\beta_k(\mu,\lambda)}^2 
&= \frac{1}{\ep{2\pi \mu}-1} \left( 1 + \ep{-\frac{\lambda \pi}{2}} \times \ep{\pi \mu}\right).
\end{split}
\end{equation}
For $\lambda \gg 1$, up to exponentially small correction, one recovers the relativistic result, i.e., the first term in the above equation. For $\lambda \ll 1$, there is an enhancement of the pair creation probability by a factor equal to $\ep{\pi \mu}$. 

Even though the asymptotic $out$ modes cannot be defined for $0 < m \leq H/2$, when $m=0$, it is again possible to define these modes since for $t\to \infty$, they behave as $e^{- i k \eta }$ where $d \eta = dt/a(t)$ is the conformal time. It is then possible to identify the massless $out$ combination of ${\rm \bf M}$ and ${\rm \bf W}$, and to extract the Bogoliubov coefficients. In this case, the norm of $\beta_k$ is
\begin{equation}
\label{betakmassless}
\begin{split}
\abs{\beta_k(\lambda)}^2 & = \frac{\pi}{\sqrt{\lambda} \abs{\Gamma\left(\frac{1}{4}+i\frac{\lambda}{4}\right)}^2 {\rm e}^{\pi \lambda /4}}\\
&+ \frac{\pi \sqrt{\lambda}}{4 \abs{\Gamma\left(\frac{3}{4}+i \frac{\lambda}{4}\right)}^2 {\rm e}^{\pi \lambda/4}}-\frac{1}{2}\, .
\end{split}
\end{equation}
For $\lambda \gg 1$, one gets $\abs {\beta_k}^2 \sim {1}/({64 \lambda^4})$, i.e., a power law decrease, unlike what we found above for the massive case.  Eq.~\eqref{betakmassless} corrects an error in Eq.~(131) of Ref.~\cite{Balbinot:2006ua} but without altering the conclusions of that section.  

\subsection{Deviations from thermality}
\label{thermidev}

Following Sec. \ref{thermbroke}, our aim is to exactly compute $T_{\rm gl}(\om,X)$ of Eq.~\eqref{tempdef} using Eq.~\eqref{Rs2}. To this end, we need to evaluate Eq.~\eqref{amp0} for quartic dispersion. Using Eq.~\eqref{chist}, we get
\begin{equation}
\label{AMP1}
\begin{split}
(\phi^{{\rm BD}, U}_{ - \om} (X) )^*  = \sqrt{\frac{\lambda}{4 \pi }} \ep{\frac{-\pi \lambda}{8}} \int_{0}^{\infty}\!\!d{p} \,  \ep{- i p H X }\,  \\
p^{ - i \frac{\omega}{H}-\frac32 } \,{{\rm \bf W}\left(\frac{-i\lambda}{4} ,\frac{i \mu }{2},\,{\frac {i p^{2}}{\lambda  }}\right)}.
\end{split}
\end{equation}
Surprisingly, it turns out that this integral can be exactly done, see Appendix~\ref{amplitude}. Since the final expression is a sum that converges as $2^{-n}$ for large $n$, one can accurately compute the ratio of Eq.~\eqref{tempdef} in terms of known hypergeometric functions. To study the consequences of quartic dispersion, we plot the temperature function $T_{\rm gl}(\om,X)$ in various cases. 

In Fig.~\ref{fig1}, we plot ${T_{\rm gl}(\omega)}/{T_H}$ as a function of $\om/H$, for various values of $\lambda$, and evaluated at $X = 0$, i.e., for an inertial detector. First, when $\om/\Lambda$ and $1/\lambda$ are both much smaller than $1$, we see that this ratio is very close to 1, as expected from former analysis~\cite{Brout:1995wp,Corley:1996ar,Balbinot:2006ua,Unruh:2004zk,Coutant:2011in}. In this robust regime, the detector will perceive a Planck law at the standard temperature, up to negligible corrections. Second, in the high frequency limit, for $\om/\Lambda \gg 1$, in agreement with the analysis of Sec.~\ref{thermbroke}, the ratio goes to $2$ irrespectively of the value of $\lambda$. This last point is not clear from the figure but can be verified analytically from the expressions of Eq.~\eqref{Asumk} and the fact that $\abs{A_\omega / A_{-\omega} } \to 1$ when $\lambda \to 0^+$. Third, we see that there is a sharp transition from the robust relativistic regime to a new regime. An examination of Eq.~\eqref{Asumk} confirms that the transition occurs at a critical frequency $\om_{\rm crit} = \Lambda/2$. 

\begin{figure}[htb] 
\includegraphics[width=8.6cm]{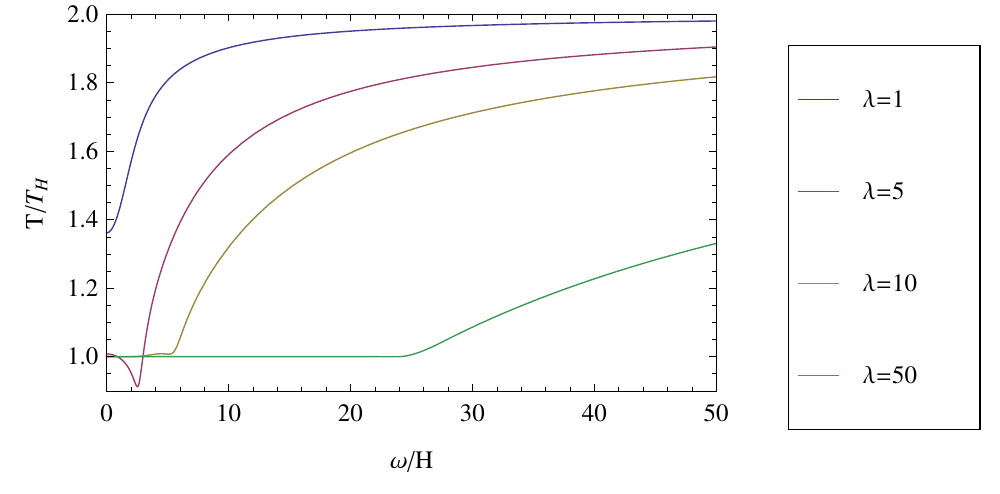} 
\caption{The ratio of $T_{\rm gl}(\om, X)$ of Eq.~\eqref{tempdef} over the standard relativistic temperature ${T_H}$ as a function of $\frac{\omega}{H}$, for $X=0$ and $m=0$, and for four values of $\lambda$, namely $1,\, 5, \, 10,$ and $50$. One clearly sees that  for large values $\lambda$, the spectrum is accurately Planckian and at the standard temperature, until $\om$ reaches a certain critical value $\om_{\rm crit}$, which is equal to $H \lambda/2$. For $\om > \om_{\rm crit}$, $T(\om,X=0)$ increases sharply and reaches $2 T_H$. This figure is essentially unchanged when we use a massive field with $\mu<\lambda/2$.} 
\label{fig1}
\end{figure} 
\begin{figure}[htb]
\centering
\includegraphics[width=8.6cm]{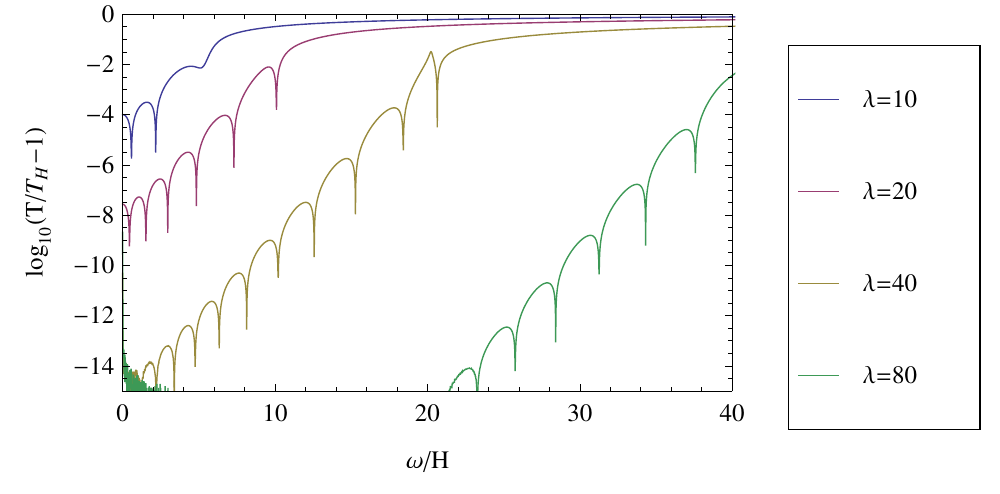}
\caption{The $\log_{10}$ of the temperature difference $ \vert T(\om) /T_H - 1 \vert $ as a function of $\omega$, for $X=0$ and $m=0$, and for four values of $\lambda$, namely $10, \, 20, \, 40,$ and $80$. We see that  $ \vert T(\om) /T_H - 1 \vert $ increases exponentially in $\om$ until $\om$ reaches $\om_{\rm crit}$. It can be shown analytically that $ \vert T(\om) /T_H - 1 \vert $ follows \eq{pexp}.} 
\label{figlog}
\end{figure}

In Fig.~\ref{figlog}, we plot $\log_{10} \vert T_{\rm gl}(\om) /T_H - 1 \vert $ to study the small deviations from the relativistic regime for $\om < \om_{\rm crit} $. We first notice that the sharp peaks are due to the fact that $T_{\rm gl}(\om) /T_H - 1$ crosses $0$ while decreasing for $\om \to 0$. A careful examination of the envelope reveals that
\be
\vert T_{\rm gl}(\om) /T_H - 1 \vert \sim \ep{-\pi \lambda/4+\pi \omega/2 H}. 
\label{pexp}
\ee
Hence, at fixed $\om$, the deviations decrease exponentially with $\lambda$, whereas, at fixed $\lambda$, they grow exponentially till $\om$ reaches $\om_{\rm crit} $. 

In Fig.~\ref{fig2} we study the $X$ dependence of $T_{\rm gl}(\om,X) /T_H$. This describes violations of the Tolman global equilibrium law. We see that the transition from the robust regime to the new regime occurs at different critical frequencies when considering detectors following different orbits labeled by $X$. Interestingly, this dependence can be expressed as
\be
\om_{\rm crit}= \frac{\lambda}{2} \left (1 - \frac{a_X}{H}  -\frac{a_X^2}{2 H^2}  +  {\cal O}\left (\frac{a_X}{H}\right )^3 \right ), 
\ee 
where $a_X =H^2 \vert X \vert/\sqrt{1-H^2 X^2} $ is the detector proper acceleration at fixed $X$. In addition, on the left panel and for $\vert HX \vert\geq 0.9$,  we notice that the low frequency temperature significantly differs from the standard one. This effect is related to the broadening of the horizon that was observed in~\cite{Finazzi:2010yq,Coutant:2011in}. In those papers, when considering perturbed metric profiles $v =v_{\rm backgrd} + \delta v$, it was found that the asymptotic black hole temperature differs from the standard one when the spatial extension across the horizon of the perturbation $\delta v$ is smaller than $\kappa x \sim (\Lambda/\kappa)^{2/3}$. Here we find that  the temperature seen by a particle detector differs from the standard one precisely when it enters this region. In a log-log plot, we have numerically found that the extension of this region (defined by the locus where the relative temperature difference is 1\%) depends on $\Lambda$ with a power equal to $0.675\pm 0.01$ in accord with the $2/3$ of the above references. Two lessons are here obtained. First, the near horizon properties can be probed either by perturbing the background metric $v$, or by introducing a local particle detector, with coherent outcomes. Second, since these responses are locally determined, they are common to de Sitter and black holes, in accord with the analysis of Appendix~\ref{BHdSc}.

\begin{figure}[htb] 
\begin{minipage}[t]{1\linewidth}
\includegraphics[width=8.6cm]{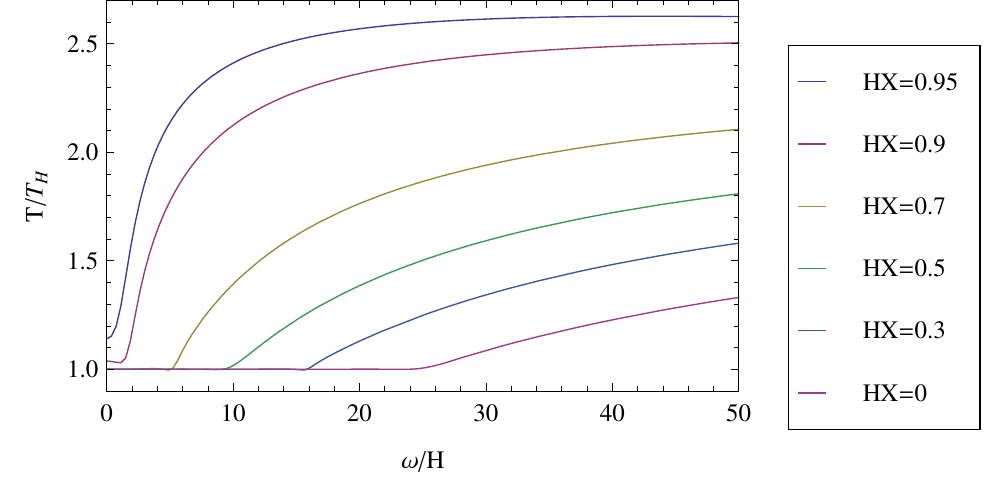}
\includegraphics[width=8.6cm]{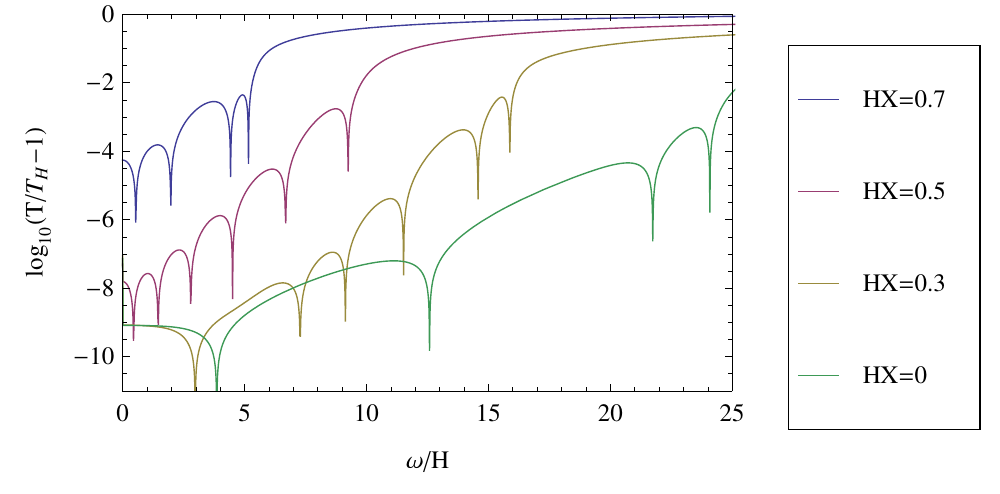}
\end{minipage}
\caption{The same ratio as in Fig.~\ref{fig1} (on the top) and Fig.~\ref{figlog} (on the bottom), for $m=0$ and $\lambda=50$, and for six different positions, namely $HX = 0, \, 0.3, \, 0.5,0.7,0.9,$ and $0.95$. On the bottom, the last two curves have not been plotted since they are too far from the other ones. The corresponding values of the acceleration of the detector are $a/H = 0,0.31 ,0.57,1,2  $, and $ 3$. One sees that ${T(\omega,X)}/{T_H}$ becomes larger than $2$ when $X \neq 0$. One also sees that the deviations at fixed $\om$ increase with $a_X$. }
\label{fig2}
\end{figure} 

Finally it is also interesting to study the behavior of $T_{\rm gl}(\om)/T_H$ when varying $\lambda$ at fixed $\om$ and for $X=0$ (see Fig.~\ref{fig5}). When $\lambda$ is large enough, i.e., larger than the critical value $\lambda_{\rm crit} = 2 \omega /H $, the deviations from the standard temperature are extremely small, in agreement to what we saw in Fig.~\ref{figlog}. Instead, for $\lambda \to 0$, $T(\om, X = 0)/T_H$ always flows to 2, with a slope that depends on the value of $\om/H$. An examination of these slopes shows that the slope decrease when $\omega$ increases: $dT/d\lambda\vert_{\lambda=0}$ goes from $1.02\pm 0.005$ to $0$. This is the behavior at small $\lambda$. The behavior at large $\lambda$ was given by Eq.~\eqref{pexp}.
\begin{figure}[!ht] 
\centering
\begin{minipage}[t]{1\linewidth}
\includegraphics[width=8.6cm]{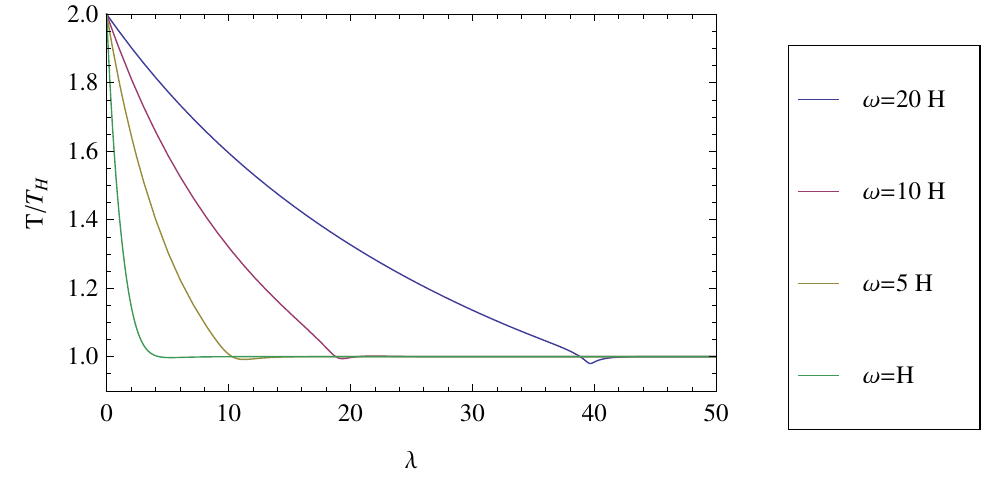}
\includegraphics[width=8.6cm]{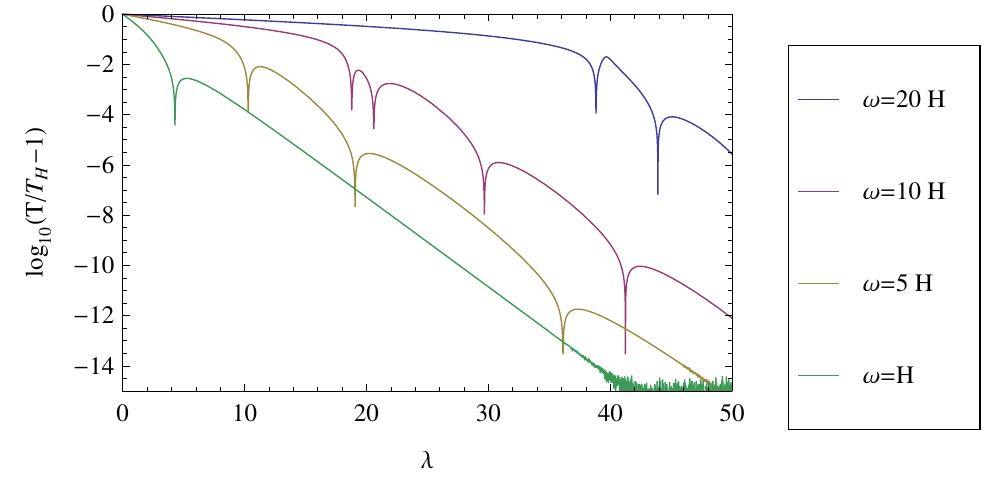}
\end{minipage}
\caption{\label{fig5} The ratio of the temperature $\frac{T(\omega)}{T_H}$ (on the top) and $\log_{10}\vert T(\om) /T_H - 1 \vert $ (on the bottom) as a function of $\lambda$, for $X=0$ and $m=0$, and for four values of $\om/H$, namely $1, \, 5, \, 10,$ and $20$. The decay of Eq.~\eqref{pexp} can be observed.}
\end{figure} 

\subsection{\texorpdfstring{Asymptotic $\boldsymbol{S}$-matrix in the $\boldsymbol{\omega}$-representation}{Asymptotic S-matrix in the omega-representation}}
\label{Smatrix}

In this section, we compute the Bogoliubov transformation between the initial BD modes and the asymptotic $out$ modes in the $\omega$ representation. In this representation, the modes are identified through their spatial asymptotic behavior, and not their temporal one we used in Sec.~\ref{betacosmo}. Hence the Bogoliubov transformation can be viewed as an $S$-matrix. This is the description which is appropriate to study the mode mixing on an analogue black hole horizon. For more details about mode identification in the $\om$-representation, we refer to Ref.~\cite{Coutant:2011in}. 

In the present case, at fixed $\om$ each basis contains $4$ modes. Hence the Bogoliubov coefficients form a $4\times4$ matrix. This matrix is an element of $U(2,2)$ since the two modes $ \phi_{\omega}^U, \,  \phi_{\omega}^V $ have a positive norm, while $ (\phi_{-\omega}^U)^*, \, (\phi_{-\omega}^V)^*$ have a negative one. In what follows we first study the massless case, and then the massive case $m  > H/2$. In both cases we shall see that the $S$-matrix possesses unusual factorization properties that are due to the two symmetries governed by $K_z$ and $K_t$. We shall also see that the elements of this matrix combine the cosmological aspects of Sec.~\ref{betacosmo} and the stationary thermal-like aspects of Sec.~\ref{thermidev}

To compute the coefficients of the $S$-matrix, we first need to identify the incoming and outgoing modes. At fixed $\om$, for quartic dispersion, the general solution of Eq.~\eqref{ommodeeqinP} contains 8 asymptotic branches, 4 for $X\to \infty$, and 4 for $X \to - \infty$. In addition, when forming wave packets in $\om$, one finds that 4 propagate towards $X = 0$, whereas $4$ propagate away from it. The mode identification is based on this second aspect: The 4 incoming modes, are, by definition, the 4 solutions that only possess one incoming asymptotic branch. These incoming modes are simply given by the Fourier transform of the stationary BD modes $\tilde \phi^{\rm BD}_\om$, thereby showing that the definitions of $in$ modes based on their temporal behavior and the spatial one are perfectly consistent.

To see this, let us consider as an example $(\phi^{{\rm BD}, U }_{- \om})^*$. Using Eq.~\eqref{ampfinal}, its asymptotic behavior can be found using~\cite{Wolfram}. Up to an irrelevant overall constant, one finds
\begin{equation}
\begin{split}
(\phi^{{\rm BD}, U}_{- \om} )^* &\underset{X\to\pm\infty}{\sim}  (1\mp 1)\times \ep{\frac{i x^2 }{2}} (\frac{i x^2 }{2})^{-i\frac{\lambda}{4}-\frac34- i \frac{\omega}{2H}} \\
+&  \left\{ Z_{\om,\lambda,\mu,\pm} \times (\frac{-i x^2 }{2})^{-\frac{1}{4} +  i\frac{\omega}{2H}-i\frac{\mu}{2}}+  (\mu \to -\mu) \right\}, 
\end{split}
\label{asmodeinom}
\end{equation}
where $x = HX \sqrt{\lambda}$ and where the coefficient $Z$ is
\be
 \begin{split}
 Z_{\om,\lambda,\mu,\pm}  &=    \frac{2^{ - i{\mu} -i\frac{\lambda}{4}+i\frac{\omega}{2H}} \Gamma(-i\mu) {\Gamma(\frac{1}{2} -  i\frac{\omega}{H}+i{\mu})  } }{ \sqrt{\pi} \Gamma(\frac12 - i\frac{\mu}{2} +i\frac{\lambda}{4}) } \\
& \times   \ep{\pm (-i \pi /4+ \pi {\mu}/{2} -  \pi{\omega}/{2H})}. 
 \end{split}
\ee
The first term in Eq.~\eqref{asmodeinom} describes the incoming high momentum branch, as can be verified by computing its group velocity $dX/dt = 1/\partial_\om P_\om$, where $P_\om=\partial_X S_\om$ is the corresponding root of Eq.~\eqref{dispHJeq}. The last two terms describe the $4$ low momentum outgoing branches. One verifies that they propagate away from the static patch, two for $X\to \infty$ and two for $X\to -\infty$. In Fig.~\ref{fig6} we schematically represent the space-time pattern associated with a wave packet made with $\phi^{{\rm BD}, U}_\om$. 
\begin{figure}[!ht] 
\centering
\label{caracteristics}
\includegraphics[width=7cm,angle=90]{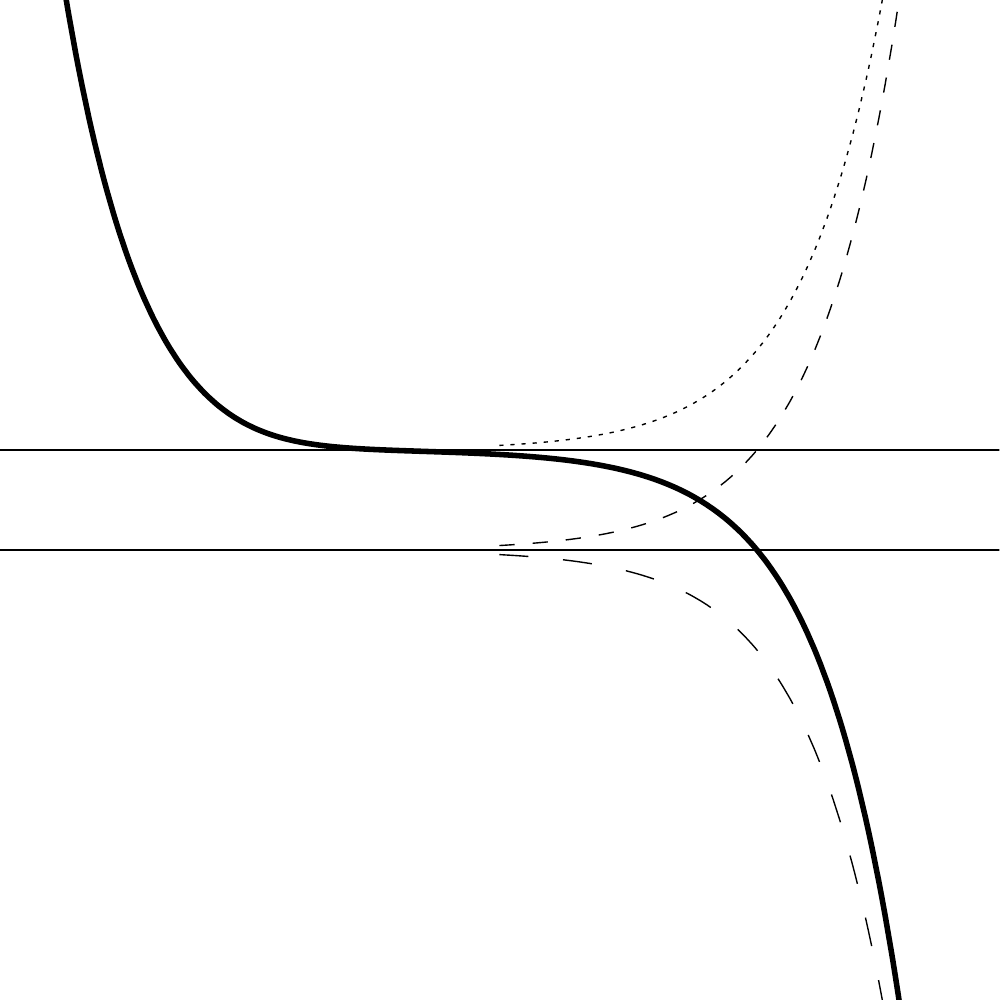}
\caption{\label{fig6}
In this figure, unlike in Fig.~\ref{penrose}, $t=$ Cst. are horizontal lines, and  $X=$ Cst. vertical ones. The two vertical lines represent the Killing horizons at $HX = \pm 1$. An incoming (BD) high momentum positive norm $U$-mode (in thick line) splits into four $out$ modes with low momenta: an outgoing  positive norm $U$-mode (thick line), a negative norm $U$-mode (dots), a positive norm $V$-mode (small dashes), and a negative norm $V$-mode (large dashes). The respective amplitudes of these four outgoing modes are given in Eq.~\eqref{modedec}.  To draw these characteristics, we work with $m=0$, $\om/H= 5$ and $\Lambda/H= 1000$. The lapse of time spent very close to the horizon is $H \Delta t \sim \ln (\Lambda/\om)$. It diverges for $\Lambda \to \infty$, in which case one recovers the relativistic behavior, and ultrahigh initial momenta.}
\end{figure} 

We now have to identify the $out$ mode basis, i.e., the four unit norm asymptotic outgoing modes. As in Sec.~\ref{betacosmo}, we treat separately the massless and the massive case. 

\subsubsection{The massless case}

Since asymptotic outgoing modes have low momentum $P$, they obey the two-dimensional d'Alembert equation. At fixed $\om$, the equation for the right moving $U$-modes is 
\be
  ( i \omega -  HX\partial_X ) \phi_\om^{U} = \partial_X \phi_\om^{U}.
\ee
For $\om > 0$, the $out$ $U$-modes of positive and negative unit norm are 
\begin{equation}
\label{phiout}
\begin{split}
 \phi_{\omega}^{U, \, out} &= \theta(1 + H X) \frac{(1 + H X)^{i\omega/H}}{\sqrt{2 \omega/H}},
 \\
 (\phi_{-\omega}^{U, \, out})^*  &=  \theta(-1 - H X) \frac{({-1 - H X })^{i\omega/H}}{\sqrt{2 \omega/H}}\, .
\end{split}
\end{equation}
The $V$-modes $ \phi_{\omega}^{V, \, out}, ( \phi_{-\omega}^{V, \, out})^*$ are obtained by replacing $X$ by $-X$ in the above. 

We put the 4 modes in a vector in the following order $\Phi_\om= (\phi_\om^U, (\phi_{-\om}^U)^*, \phi_\om^V, (\phi_{-\om}^V)^* )$, both for the BD and the $out$ modes, and we define the $S$-matrix by $\Phi^{BD}_\om = S_\om \Phi^{out}_\om$. We find that $S_\om$ factorizes as
\begin{equation}
\begin{split}
S_\om= \left[ \begin{array}{cccc}
\alpha_k &0 &0 &\beta_k\\
0& \alpha_k ^*&\beta_k^*  &0\\
0 &\beta_k &\alpha_k  &0\\
\beta_k^* &0 &0 &\alpha_k ^* \\
\end{array}\right] \times 
\left[ \begin{array}{cccc}
\alpha_\omega^H &\beta_\omega^H& 0&0\\
\beta_\omega^H&\alpha_\omega^H &0&0\\
0&0&\alpha_\omega^H &\beta_\omega^H\\
0&0&\beta_\omega^H &\alpha_\omega^H\\
\end{array}\right] .
\label{Smassl}
\end{split}
\end{equation}
Moreover, the Bogoliubov coefficients $\alpha_k, \beta_k$ are those of Eq~\eqref{betakmassless}, and $\alpha_\om^H, \beta_\om^H$ are the standard relativistic coefficients, taken real, and obeying $\beta_\om^H/\alpha_\om^H = e^{- \pi \om/H} $ and $\vert \alpha_\om^H\vert ^2 - \vert \beta_\om^H \vert ^2= 1 $. To get these real coefficients we chose the (arbitrary) phases of the $out$ modes in an appropriate manner. There are only 4 different coefficients in $S_\om$, and they all have a clear meaning when considering one BD mode. In fig.~\ref{fig6}, we represent the mode 
\begin{equation}
\begin{split}
\phi^{{\rm BD}, \, U}_\omega &= \alpha_\om \,  \phi^{out, \, U}_{\omega} + \beta_\om \, (\phi^{out, \, U}_{-\omega})^* \\
&+ A_\om \, \phi^{out, \, V}_{\omega} + B_\om \, (\phi^{out, \, V}_{-\omega})^* .
\end{split}
\label{modedec} 
\end{equation}
The $\alpha_\om, \beta_\om$ coefficients weigh the mode mixing amongst $U$-modes of opposite norm, whereas  $A_\om$ and $B_\om$ describe respectively the elastic and the anomalous $U-V$ mode mixing. The norm of these four coefficients obey 
\begin{equation}
\begin{array}{rlrl} 
\abs {\alpha_\om} ^2 &= \abs{\alpha_k}^2 \times (n^H_\om + 1), &
\abs {\beta_\om} ^2 &= \abs{\alpha_k}^2 \times n^H_\om , \\
\abs {A_\om} ^2 &= \abs{\beta_k}^2 \times  n^H_\om , &
\abs {B_\om} ^2 &= \abs{\beta_k}^2 \times  (n^H_\om +1) ,
\end{array}
\end{equation}
where $n^H_\om = 1/(e^{\om/T_H} - 1)$ is the Planck spectrum at the standard temperature $T_H$. We see that the deviations from the relativistic spectrum are proportional to $\abs{\alpha_k}^2 - 1 = \abs{\beta_k}^2 \sim \lambda^{-4}$, as those breaking the relativistic $U$-$V$ decoupling. Thus both deviations from the relativistic theory are governed the cosmological pair creation rates at fixed $\bf k$. We notice that the decay of the deviations from thermality in $1/\lambda^4$ is in agreement with the decay in $1/\omega_{\rm max}^4$ found in a black hole metric when working at fixed $D$, see Fig.~14 of Ref.~\cite{Macher:2009tw}. We also notice that irrespectively of $\om$ and $\Lambda$, the elastic $\abs {A_\om} $ is the smallest coefficient. We finally emphasize that these extremely simple results are exact, and follow from the hypergeometric functions $\ _2F _2$ of Eq.~\eqref{ampfinal}. 

\subsubsection{The massive case}

As in Sec.~\ref{betacosmo}, the  massive $out$ modes should be handled with care. An orthonormal basis for these $out$ modes is given by the following right modes $(R)$:
\begin{equation}
\label{phioutmassif}
\begin{split}
\phi^{out}_{R,\omega}= \theta(X) \frac{(HX)^{i\frac{\omega}{H}- i\mu}}{\sqrt{2 \mu  H X}},
\quad
\phi^{out\,*}_{R,-\omega}= \theta(X) \frac{(HX)^{i\frac{\omega}{H}+ i\mu}}{\sqrt{2 \mu  (H X)}},
\end{split}
\end{equation} 
together with the $L$-modes obtained by replacing $X$ by $-X$ in the above expressions. We have used this $R$-$L$ separation in the place of the $U$-$V$ one based on the sign of the group velocity, because, for massive modes the asymptotic group velocity with respect to the flow $v = HX$ is no longer well defined. 

We now put the 4 $out$ modes in a vector in the following order $\Phi_\om= (\phi_\om^R , (\phi_{-\om}^R)^* , \phi_\om^L,(\phi_{-\om}^L)^* )$, while the 4 $in$ modes are ordered in the same order as in the massless case. Defining again the $S$-matrix by $\Phi^{BD}_\om = S_\om \Phi^{out}_\om$, we obtain 
\begin{equation}
\begin{split}
S_\om= \left[ \begin{array}{cccc}
\alpha_k &0 &0 &\beta_k\\
0& \alpha_k ^*&\beta_k^*  &0\\
0 &\beta_k &\alpha_k  &0\\
\beta_k^* &0 &0 &\alpha_k ^* \\
\end{array}\right] \times 
\left[ \begin{array}{cccc}
T_\om &0& R_\om&0\\
0&T_{-\om} &0&R_{-\om}\\
R_{\om}&0&T_\om &0\\
0&R_{-\om}&0&T_{-\om} \\
\end{array}\right] .
\end{split}
\end{equation}
On the left matrix, the $\alpha_k, \beta_k$  coefficients are those of Eq~\eqref{betakmassive}. Hence, as far as this matrix is concerned, we obtain the same structure as in Eq~\eqref{Smassl}. Instead on the $\om$-dependent right matrix, the coefficients are 
\begin{equation}
T_{\om}=\frac{\ep{ - \pi(\mu - \omega/H)/2 }}{\sqrt{2\cosh \pi( \mu - \omega/H) }},
\ 
R_{\om}=\frac{\ep{\pi(\mu - \omega/H )/2 }}{\sqrt{2\cosh \pi( \mu - \omega/H) }}. 
\end{equation}
They obey $\abs{T_{\om}}^2 + \abs{R_{\om}}^2 = 1$. Hence unlike what was found in Eq.~\eqref{Smassl} the right matrix now describes an elastic scattering between modes of the same norm. As a result, the main difference between the massless and the massive case is that the final occupation number of massive particle no longer diverge as $T_H/\om$ for $\om \to 0$. This disappearance of the thermal like divergence was already found in Ref.~\cite{Coutant:2012zh} in black hole metrics.

\section{Conclusions}

In this paper we obtain three kinds of results, precise mathematical ones characterizing dispersive fields in de Sitter space, those concerning the correspondence between dispersive effects in de Sitter and for black holes, and finally more general ones associated with the observation that thermality is violated when Lorentz invariance is broken at high energy. 

Concerning the first kind, in Sec.~\ref{setting}, we used the group associated with the two residual symmetries of dispersive fields in de Sitter to provide precise relationships between the two representations of the field, based respectively on the homogeneity and on the stationarity of the settings. The key result is that the homogeneous modes and the stationary ones can be {\it all} expressed in terms of the single BD mode $\chi_{\rm BD}(P)$ and its complex conjugated, where $\chi_{\rm BD}$ obeys \eq{Pmodeeq} and \eq{chiWKB}, see Eqs.~\eq{kp} and~\eq{omegaP}. For free fields, all observables are thus encoded in that single mode. The algebraic properties associated with the residual group are further explored in Appendix~\ref{group} and shown to be compatible with both dispersive and dissipative effects.

Having identified this group, we present in Appendix~\ref{BHdSc} the precise correspondence between high frequency dispersion in de Sitter and in black hole backgrounds. Because the de Sitter case is also stationary, many aspects are common to both cases, with one exception. In de Sitter, when the preferred frame coincides with the cosmological frame, the fields $u, \, s$  obey the affine algebra of Eq.~\eqref{usc}. Instead, in stationary black hole space-times endowed with a freely falling frame, $u$ and $s$ obey the {\it local} algebra of Eq.~\eqref{commus} governed by $\Theta(x)$, the expansion of $u$. Since this is basically the only difference, the observables of dispersive fields computed in black hole backgrounds, such as the $S$-matrix, possess the same properties as in de Sitter, up to inverse powers of $D^{3/2}\Lambda/\kappa$, where $\Lambda$ is the dispersive frequency, $\kappa= \Theta_0$ is the expansion evaluated on the horizon, $D= \kappa x$ gives the extension of the near horizon black hole region which can be mapped onto de Sitter, and where the power $3/2$ characterizes quartic dispersion~\cite{Coutant:2011in}. As indicated in footnote~\ref{gamma}, this correspondence possesses a generalization when the preferred frame is not freely falling.

In Sec.~\ref{thermal} we show that the two-point function computed in the BD vacuum is still stationary and periodic in Im$t$ with period $2\pi/H$, as it is for Lorentz invariant fields. In spite of this, we then show that the BD vacuum is no longer a thermal state when restricted to the static patch. In particular, we show that the temperature function of Eq.~\eqref{tempdef} is, for ultrahigh frequency $\om /\Lambda \gg 1$, $n$ times the standard one, where $n$ is the highest power of $P^2$ in the dispersion relation of \eq{disprel}. In Sec.~\ref{thermidev}, by considering the response function of particle detectors with different acceleration, we also show that the Tolman law is violated. Even though the BD vacuum is no longer in thermal equilibrium, we prove that (for free fields at least) it is still the only stationary, regular, and stable state, as it is in relativistic theories~\cite{Marolf:2010nz,Marolf:2011sh,Hollands:2010pr}. In other words, for dispersive fields, there is no (regular) KMS state on de Sitter space. We believe that these properties will remain true when considering interacting fields. Finally we {explain} the origin of the violations of thermality in terms of the loss of the positivity of the stationary Hamiltonian restricted to the static patch. Whereas this operator possesses a spectrum bounded from below for Lorentz invariant theories, this is no longer true for dispersive fields. As a result the ordinary second law of thermodynamics is no longer protected, violations of this law are possible, and the system might develop dynamical instabilities.

In this respect the fact that the BD vacuum is shown to be stable in de Sitter becomes a nontrivial result. Moreover, because of the precise correspondence between dispersive effects in de Sitter and in black hole metrics, we conjecture that the properties found in de Sitter should apply to black holes, when these are dynamically stable, or sufficiently long living, in the adopted theory of gravity that violates Lorentz invariance at high energy.

\begin{acknowledgments}
{We are grateful for many interesting discussions with Julian Adamek, Antonin Coutant, Ted Jacobson, Stefano Liberati, Julien Serreau and Robin Zegers. R.P. thanks the organizers of the 17th Conference of Cosmology in Peyresq in June 2012 for the opportunity to present and discuss this work.} 
\end{acknowledgments}

\appendix

\section{Affine subgroup and ultraviolet dispersion}
\label{group}

In this Appendix we explore the relationships between the residual affine group based on the two generators $K_t, K_z$ of Eq.~\eqref{alg}, and the possibility of considering modifications of the field equation that encode dispersive and/or dissipative effects. In a sense we are performing a group theoretical approach to dispersion and dissipation on de Sitter space. For a similar approach based on the full de Sitter group, we refer to Ref.~\cite{Joung:2006gj}.

The three generators of $SO(1,2)$ can be taken to be 
\begin{equation}
\begin{split}
K_t &=\partial_t - H z \partial_z , \quad
K_z=K_-= \partial_z,  \\
K_+&= -H z \partial_t +\left( \frac{z^2 H^2+\ep{-2H t}}{2} \right)\partial_z.
\end{split}
\end{equation}
They are linked to the usual generators by $K_\pm= K_x \pm K_y$, and they obey 
\begin{equation}
\left[K_t,K_z\right]= H K_z, \ 
\left[K_t,K_+\right]= -H K_+, \ 
\left[K_+,K_z\right]= H K_t .
\end{equation}
We now wish to characterize the set of local differential operators that commute with both $K_t$ and $K_z$. The most general local operator acting on scalar fields can be written as
\be 
\hat O = \sum_{n,m=0}^\infty \alpha_{n,m}(t,z)\,  \partial_z^n \, \partial_t^m. 
\ee 
Imposing that $\hat O$ commutes with $K_z$, implies that the $\alpha$'s depend only on $t$. Imposing that it also commutes with $K_t$ implies $\alpha_{n,m}(t)=\alpha_{n,m} \, \ep{- n Ht}$, where $\alpha_{n,m}$ are constants. Hence, $\hat O$ is necessarily of the form
\be
\hat O = \sum_{n,m=0}^\infty \alpha_{n,m}\, \hat  P^n\,  \hat \Omega^m
\ee
where $\hat P = -i \ep{-H t}\partial_z = -i s^\mu \partial_\mu$ is the preferred momentum operator, and $\hat \Omega =  i\partial_t = - i u^\mu \partial_\mu$ is the preferred frequency entering in Eq.~\eqref{disprel}. What we learned here is that the only vector fields that commute with $K_t$ and $K_z$ are the $u$ and $s$ fields associated with the cosmological frame. In addition we notice that $u$ and $s$ obey 
\be
\left[u,s\right] = H s ,
\label{usc}
\ee 
which is the affine algebra of Eq.~\eqref{alg}. Hence this algebra is intrinsic to the cosmological frame on de Sitter space.~\footnote{
Equation~(\ref{usc}) follows from the fact that the commutator $[u,s]$ must be a linear combination of $u$ and $s$ since they are the only fields that commute with $K_t$ and $K_z$. It is equal to $H s$ because $u$ has been chosen to be freely falling, see footnote~\ref{gamma} for the general case. 
} 
Notice also that Eq.~\eqref{usc} is exact only in de Sitter. However, an interesting generalization of this equation exists in black hole spaces, see Eq.~\eqref{commus}.

Our program is to treat $\hat O$ as defining the field equation that generalizes Eq.~\eqref{dispKGeq}. To this end, we impose that $\hat O$ be second order in $ \hat \Omega$. This leads to 
\be
\hat O = - \hat \Omega^2 + g(\hat P) \, \hat \Omega  +h(\hat P).
\label{Ook}
\ee
Next we impose the invariance under the discrete parity symmetry  $z\rightarrow-z$. This implies that $g$ and $h$ are even functions of $\hat P$. In higher dimensions, this condition would follow from the requirement of isotropy. The last important condition is that $\hat O$ be compatible with a unitary evolution~\cite{Parentani:2007uq}. The proper way to specify this condition is the following: the part of $\hat O$ that is even in $\hat \Omega$ describes dispersive effects and should be self-adjoint, whereas the odd part describes dissipative effects and should be anti-self-adjoint, where the adjoint is defined by
\be
\begin{split}
 \int d^2x \sqrt{-g} \,  \Phi^* \, (\hat O \Psi ) &= \int d^2x \sqrt{-g} \, (\hat O^\dagger \Phi)^* \, \Psi .
\end{split}
\ee 
To sort out the contributions which are due to the expansion, is useful to introduce the self-adjoint operators $\hat \Omega_{\rm sa} = \frac12\left (\hat \Omega + \hat \Omega^\dagger\right )= \hat \Omega + i \frac{H}{2}$, and $\hat \Omega_{\rm sa}^2 = \hat \Omega^2 +i  H \hat \Omega - \frac{H^2}{4}$. Then, the \enquote{unitary} operators are given by
\begin{equation}
\begin{split}
\label{hatOf}
\hat O
 &= -\hat \Omega_{\rm sa}^2 -  i \left ( \gamma_{sa}  \hat \Omega_{\rm sa} + \hat \Omega_{\rm sa} \gamma_{sa}  \right ) + F_{sa},
\end{split}
\end{equation}
where $\gamma_{sa} $ and $F_{sa}$ are both real functions of $\hat P^2$. To be more explicit, when applied to $e^{i {\bf k}z}\phi_{k}(t) $, the field equation $\hat O \Phi = 0$ gives
\be
\begin{split}
&\left( \frac{1}{a}\, \partial_t \, a  \partial_t + 2\, (\gamma_{sa}  a)^{1/2} \frac{1}{a}\partial_t (\gamma_{sa}  a)^{1/2}\right .\\
& \hspace{3cm}+\left .  ( F_{sa} +   \frac{H^2}{4}) \right) \phi_{{k}}(t) =0 , 
\end{split}
\ee
where $a$ is the scale factor, and where the argument of $\gamma_{sa} $ and $F_{sa} $ is $k^2/a(t)^2$.  Because of the affine group, this equation can be simplified using the function $\chi(P)=\phi_{k}(t)$ introduced in Sec~\ref{Mode analysis}. In the present more general case, one still gets a single equation valid for all $\phi_{\bf k}$ modes:
\be
\begin{split}
&\left(H^2 P^2 \partial_P^2 - (\frac{\gamma_{sa} }{ P} )^{1/2} H P^2 \partial_P (\frac{\gamma_{sa} }{ P})^{1/2}   \right .\\
& \hspace{3cm} \left . +  ( F_{sa} +   \frac{H^2}{4}) \right)\chi = 0.
\end{split}
\ee
The function $F = F_{sa} + \frac{H^2}{4}$ describes the dispersion effects compatible with the affine group, exactly as in Eq.~\eqref{kmodeq}. Instead, the function $\gamma_{sa} $, which multiplies the odd term in $\Omega$, describes the dissipative effects compatible it. It precisely matches the set of $\gamma$ functions introduced in~\cite{Parentani:2007uq,Adamek:2008mp} to describe dissipative effects that are local in time, and that obey the generalized equivalence principle, which states that the action must be a sum of scalars under general coordinate transformations which reproduce those one had in Minkowski space-time endowed with a homogeneous static $u$ field. This agreement is nontrivial and follows from the fact that, on one side, the GEP implies that the field equation can only depend on the two {\it scalars} $\hat \Omega$ and $\hat P$ defined by the metric $g$ and the $u$ field, whereas, on the other side, $\hat \Omega$ and $\hat P$ are the two {\it invariants} under the generators $K_t$ and $K_z$ of the affine group.  

It is interesting to impose the invariance under the third generator, namely $K_+$. In that case, the only invariant operator is $\hat O = -\hat \Omega_{sa}^2 + \hat P^2  =   - K_t^2 + K_+ K_- + K_- K_+ $ which is the Casimir of $SO(1,2)$. We thus see that neither dispersion nor dissipation is compatible with the full de Sitter group. We also notice that the affine group has no Casimir operator in the sense that the universal enveloping algebra of the affine group has no element, but the identity, that commute with the affine group.

As a final comment, we notice that the affine group is closely related to Fourier and Mellin analysis~\cite{morse1953methods}. When working  on $\mathcal{L}^2(\mathbb{R})$, the eigenmodes of $-i K_z = -i \partial_z$ of eigenvalue $k$ are the plane waves $e^{ikz}$,  whereas those of $-iK_t=-i H (z\partial_z+1/2)$ of eigenvalue $\om$, are $\phi_\om^\pm = \theta(\pm z)(\pm z)^{i \om/H-1/2}$. The latter live on either side of $z= 0$ and they correspond to Mellin modes. They are complete for $\om \in \mathbb{R}$ (since invertible) on $\mathcal{L}^2(\mathbb{R}^+)$. Hence, to have completeness on functions of $\mathbb R$, one need two families of Mellin modes, on either side of $z=0$, given by $\phi_\om^\pm$. 

\section{Black hole--de Sitter correspondence} 
\label{BHdSc}

Let us start afresh with a stationary black hole metric and a preferred frame that we describe by a unit time-like vector field $u$. We also introduce the unit space-like vector field $s$ orthogonal to $u$. To perform the comparison with the de Sitter case in meaningful terms we shall use quantities, and coordinates, that are invariantly defined.
 
To get a situation which is closer to that of Sec.~\ref{dSspace}, we now add two assumptions. We first assume that the preferred frame is freely falling, i.e., $\gamma^\nu \doteq u^\mu D_\mu u^\nu =0$. This implies that the commutator of $u$ and $s$ obeys
\be
[u,s] = \Theta\,  s ,
\label{commus}
\ee
where $\Theta(x^\mu) \doteq - D_\mu u^\mu$ is the expansion of the $u$ field. This important equation generalizes Eq.~\eqref{usc}.  [To obtain it, we used $g=-uu+ss$ which implies $-D_\mu u_\nu = u_\mu \gamma_\nu + \Theta s_\mu s_\nu $, and the Lie derivative $\mathcal{L}_u (s^\mu\otimes s^\nu) = \mathcal{L}_u (g^{\mu\nu}) =  - D^\mu u^\nu - D^\nu u^\mu = u^\mu \gamma^\nu+u^\nu \gamma^\mu + 2\Theta s^\mu s^\nu$. Hence, $[u,s] = \Theta\,  s + \gamma\, u$ where $\gamma =  \sqrt{\gamma^\mu \gamma_\mu}$.] We then assume that $u$ commutes with  the stationary Killing field $K_\tau$. Under these assumptions, when using the preferred frame coordinates $d\tau \doteq u_\mu dx^\mu$, $\partial_x \doteq s^\mu \partial_\mu$, the metric reads
\be
\label{metricBH}
ds^2 = - d\tau^2 + (dx - v(x) d\tau)^2, 
\ee
the expansion of $u$ is $\Theta = \partial_x v$, and the norm of $K_\tau= - \partial_\tau\vert_x$ is $K_\tau^2 = - 1 + v^2$. The location of the Killing horizon, where $K_\tau^2 = 0$, is taken to be $x=0$. Then the behavior of $v$ in near horizon region (NHR) is  $v \sim - 1 + \kappa x$, where $\kappa = \Theta_0$ is the expansion evaluated on the horizon~\cite{Jacobson:2008cx}. When $\kappa > 0$ and $v < 0$, one has a black hole horizon, since null outgoing geodesics follow $x \sim x_0 \, e^{\kappa \tau}$ in the NHR.

The important lesson here is that under the assumptions of stationarity and freely fallingness, the black hole metric and the preferred frame are completely, and invariantly, determined by $v(x)$. Since the de Sitter background fields of Sec.~\ref{dSspace} can be described by the same settings with the extra condition that $v_{dS}$ is linear in $x$, the comparison of dispersive effects associated with Eq.~\eqref{disprel} can be easily done for the solutions of both Eq.~\eqref{dispHJeq} and Eq.~\eqref{dispKGeq}. In particular, we can already predict that the {\it deviations} between de Sitter and the black hole case will be governed by the spatial extension of the black hole NHR where $v$ is approximatively linear in $x$. When $v= -1 + D \tanh(\kappa x/D)$, the extension is, roughly speaking, given by $\vert \kappa x \vert = D$.~\footnote{
Even though the parameter $D$ plays no role when computing the Hawking spectrum using relativistic fields, it plays important roles in black hole physics. First, the deviations with respect to the Hawking spectrum due to dispersion are governed by $D$~\cite{Macher:2009tw,Finazzi:2010yq,Coutant:2011in}. Second, the nonlocal correlations across a black hole horizon are also governed by $D$, in that they start to differ from vacuum correlations when $\kappa x \sim D$~\cite{Schutzhold:2010ig,Parentani:2010bn}}
Using this velocity profile, near the horizon,  Eq.~\eqref{commus} is given by
\be
[u,s] = \Theta(x)\,  s=  \kappa \, s \, \left(1- \frac{(\kappa x)^2 }{ D^2} + O(\kappa x)^4\right) ,
\label{usnhr}
\ee
which clearly shows that the deviations with respect to Eq.~\eqref{usc} are governed by $\kappa x/D$. 

We now wish to show  that this correspondence is not limited to the background fields, but extends to the {\it dynamics} of dispersive fields. At the classical level, this is most clearly seen by considering Hamilton's equations. In particular, irrespectively of the choice of $f$ in Eq.~\eqref{disprel}, the time derivative of the momentum $p=\partial_x S = s^\mu \partial_\mu S$ obeys
\be
\frac{dp}{d\tau} = - \frac{1}{\partial_\om x_\om(p)} = -p \, {\Theta(x_\om(p))}\, ,
\label{ptau}
\ee
where $x_\om(p)$ is the root of Eq.~\eqref{disprel} at fixed $\om$, i.e., with $\Omega$ expressed as $\Om = \om - v(X) P$. We learn here that Eq.~\eqref{ptau} is the dynamical equivalent of Eq.~\eqref{commus}. This establishes how the preferred frame algebra imprints the particle's dynamics. Having understood that, as long as $\kappa x \ll D$, Eq.~\eqref{ptau} and Eq.~\eqref{usnhr} guarantee that $p$ obeys 
\be
p(\tau) = p_0 \, e^{- \kappa \tau}, \,
\ee
 as in the de Sitter cosmology where $P = k/a(t)$. It is worth pointing out that this exponential redshift applies for both signs of $p$, i.e., for both right and left moving solutions. This correspondence in $p$-space also applies to the classical trajectories in $x$-space. At fixed $\om$, $x(\tau)$ obeys $\om - v(x) p= \pm F(p)$, where $p(\tau)$ is the solution Eq.~\eqref{ptau}, and where $+$ ($-$) describes right moving trajectories. As long as $v \sim -1 + \kappa x$ furnishes a good description of $v$, the dispersive trajectories $x_\om(\tau)$ in the black hole metric are indistinguishable from those in de Sitter, i.e., $x_\om(\tau)$ is the same function as $X^{dS}_\om(t) - 1/H$ for $H = \kappa$ and $t = \tau$.~\footnote{
\label{gamma}
When the preferred frame is not freely falling, as found in extended theories of gravity~\cite{Eling:2006ec,Blas:2011ni,Barausse:2011pu,Berglund:2012bu} and in analogue gravity, there exists an interesting generalization. Eqs.~(\ref{commus}, \ref{metricBH}) are replaced by $[u,s] = \Theta\,  s + \gamma u $, and $ds^2 = -c^2 d\tau^2 + (dx - v d\tau)^2$, where $x$ is still defined by $s^\mu \partial_\mu = \partial_x$. When $[u, K_\tau] = 0$, one has $\partial_x v = c \Theta$ and $\partial_x c = c \gamma$. The acceleration $\gamma$ is thus described by what plays the role of a varying speed of light $c(x)$. Using $c$, we get
\begin{equation}
\kappa = \partial_x (c+v)\mid_{x=0} = c_0 (\Theta_0 + \gamma_0), 
\end{equation}
thereby recovering the expression of the surface gravity used in condensed matter models~\cite{Barcelo:2005fc,Macher:2009nz}, and generalizing~\cite{Jacobson:2008cx}. Moreover, Eq.~(\ref{ptau}) becomes  ${dp}/{d\tau}= -c( \Theta p +  \gamma F(p))$. In the NHR, this gives  ${dp}/{d\tau}= - \kappa p -  c_0 \gamma_0 (F(p)-p)$. Using the techniques of Ref.~\cite{Coutant:2011in}, one finds no spectral deviation at first order in $\gamma_0$ for sufficiently low $\omega$. Finally, the black hole--de Sitter correspondence is maintained when considering a preferred frame in de Sitter whose acceleration matches $\gamma_0$. The new field $u_{\gamma_0}$ is related to those of \eq{usc} by a boost: $u_{\gamma_0} = u \cosh \zeta  + s \sinh\zeta $, where $\gamma_0 = H\sinh \zeta$. 
In this case, there is a "universal horizon"~\cite{Blas:2011ni,Berglund:2012bu} at $HX = \coth \zeta$, as in black hole metrics. We are planning to study the spectral consequences of $\gamma_0$ in a forthcoming publication.}

The correspondence further extends at the level of the dispersive field because the stationary modes $\phi_\om$ still (exactly) obey Eq.~\eqref{ommodeeq} in the black hole case. Therefore, near the Killing horizon, the black hole Fourier modes $\tilde \phi_\om(p)$ factorize as in Eq.~\eqref{facto}, where $\chi$ will obey Eq.~\eqref{Pmodeeq} with $H = \kappa$. At this point we make two observations. First, Eq.~\eqref{Pmodeeq} resulted in de Sitter from the coexistence of $K_t$ and $K_z$, and their algebra of Eq.~\eqref{alg}. Second, Eq.~\eqref{Pmodeeq} was used in all analytical treatments of the scattering of dispersive modes on a black hole horizon~\cite{Brout:1995wp,Corley:1996ar,Balbinot:2006ua,Unruh:2004zk,Coutant:2011in}. These observations raise several questions: 
\begin{itemize}

\item What is the relevance of this correspondence for the $S$ matrix ? 

\item What is the validity domain of this correspondence in terms of time lapses ?

\item Can we define a field $K_z$ which is approximatively Killing near the horizon ? 

\end{itemize}
The first question is certainly the most important one. As shown in Refs.~\cite{Coutant:2011in,Finazzi:2010yq,Finazzi:2012iu}, in the black hole case, when $\Lambda/\kappa \gg 1$, the {\it leading deviations} from the Planck spectrum at the standard Hawking temperature are governed by inverse powers of the parameter $D$ which enters in Eq.~\eqref{usnhr}. This means that these deviations are in fact defined with respect to the corresponding dispersive spectrum evaluated in de Sitter space. This is perfectly coherent because in de Sitter, the deviations due to dispersion with respect to the relativistic spectrum are very small, see Sec.~\ref{betacosmo} and Sec.~\ref{Smatrix}, much smaller than those of the black hole case. In brief, this explains why the parameter $D$ of Eq.~\eqref{usnhr}, which governs the extension of the black hole near horizon region which can be mapped in de Sitter, also governs the spectral deviations of the black hole flux. 

Concerning the second question, as far as space is concerned, the validity range of the linearized expression of $v$ around $K_\tau^2 = 0$ is trivially fixed by $D$. What is less trivial concerns the lapse of time during which this linearized expression can be used, given the dispersion relation of Eq.~\eqref{disprel}. It is at this level that the separation between the background scale $\kappa = \Theta_0$ and the dispersive scale $\Lambda$ enters. When $\Lambda/\kappa \gg 1 $, the lapse of time during which the right moving $U$-particles of frequency $\om \sim \kappa$ stay in the NHR scales, for quartic dispersion, as $\kappa \Delta \tau \sim \log(D^{3/2}\Lambda/\kappa)$. Correspondingly, the accumulated redshift from the high initial momentum till the final one scales as $p_{in}/p_{out} \sim e^{\kappa \Delta \tau }\sim D^{3/2}\Lambda/\kappa$. We see that it combines in a nontrivial manner the scale separation and the spatial extension of the NHR. In Ref.~\cite{Finazzi:2012iu} it was explicitly shown that $\kappa \Delta \tau $, the adimensional lapse of time spent in the de Sitter like region, governs the  properties of the black hole spectrum.

Having clarified these issues, it is worth returning to geometrical aspects by investigating how a vector field $K_z = \partial_z$ can be introduced in black hole space-times and to what extent it could be considered as an \enquote{approximate Killing field}. It should be first pointed out that, a priori, there exist several ways to introduce a new coordinate $z$. Indeed, in de Sitter, $K^{dS}_z$ obeys several properties that can be used to define the vector field in the black hole case. For instance, the commutator $[u,K^{dS}_z]$ vanishes. Using this property to define $z$, one gets the construction of Ref.~\cite{Parentani:2007uq} where the black hole metric reads $ds^2 = - d\tau^2 + a^2 dz^2$, with $a = v(x(\tau,z))/v(z) \sim e^{-\kappa \tau}$ in the NHR. The disadvantage of this choice is that the lapse of time during which the exponential is found is much shorter than the lapse $\Delta \tau$ we above discussed. \textit{A posteriori}, it turns out that a better choice is provided by imposing that
Eq.~\eqref{alg} be satisfied:
\be
[K_\tau, K_z] = \kappa \, K_z \, .
\label{2KT}
\ee
This implies that $K_z \doteq  \ep{ \kappa t} \partial_x $ is the derivative with respect to the new coordinate $z=x \ep{-\kappa t}$. We then have the following commutation relations 
\begin{equation}
[K_z,u] = (\Theta(x) -\kappa)\,  K_z\, ,
\end{equation}
and 
\begin{equation}
D_\mu K_{z\, \nu} +D_\nu K_{z\, \mu} = \frac{\Theta(x)-\kappa}{2} (s_\mu u_\nu + u_\mu s_\nu) \, .
\end{equation}
Since $\kappa = \Theta(x=0)$, we see that the deviations from the Killingness, i.e., the second equation, and from the homogeneous de Sitterness, the first equation, are both governed by the gradient of $\Theta$ in the NHR, and not from $\Theta_0 = \kappa$ itself. It is thus geometrically meaningful, and dynamically relevant, to say that a stationary black hole metric endowed with a freely falling frame possesses, in the NHR, an approximate homogeneous Killing field $K_z$ obeying the affine algebra of Eq.~\eqref{2KT}. 

\section{Evaluation of Eq.~\texorpdfstring{\pmb{\eqref{AMP1}}}{}} 
\label{amplitude}

To obtain an explicit expression for $(\phi^{{\rm BD}, U}_{- \om})^*$  of Eq.~\eqref{AMP1}, we shall compute a more general function $A(z)$ to be able to exploit some analytical property in $z$. It is given by
\begin{equation}
A(z)\! \doteq\! \ep{i\pi(\alpha+1)/4} \int_{0}^{\infty} \! \! d{p}\,  p^{\alpha} \,  \ep{- i p x } \, {\rm \bf W}\left( \kappa, \nu,\,{{i z p^{2}} }\right) \frac{\ep{i(z-1)p^2}}{  z^{\frac12}}, 
\ee
and which is related to the BD mode by
\be
(\phi^{{\rm BD}, U}_{ - \om}(X))^* = \ep{-i\pi(\alpha+1)/4} \sqrt{\frac{\lambda}{4 \pi }} \ep{\frac{-\pi \lambda}{8}} \lambda^{(\alpha+1)/2} A(z =1).
\end{equation}
To simplify notations, we introduced $\alpha=-3/2 - i {\omega}/{H}$, $\xi = {-i\lambda}/{4} $, $\nu ={i \mu }/{2}$, $x=H X\sqrt{\lambda}$ and rescaled $p\to p\sqrt{\lambda}$. Making a rotation in complex $p$ plane of angle $\pi/4$, one gets
\begin{equation}
\begin{split}
A(z)&=\int_{0}^{\infty}d{p} p^{\alpha} \ep{- p x\ep{i\pi/4} }\, {\rm \bf W}\left( \xi, \nu,\,{z p^{2}}\right) \, \frac{\ep{(z-1)p^2}}{  z^{\frac12}}.
\end{split}
\end{equation}
The Whittaker is then expressed as a sum ${\rm \bf W}\left( \xi, \nu,\,{z p^{2}}\right)= B(\nu)+B(-\nu)$ where~\cite{Abramowitz} 
\begin{equation}
\begin{split}
B(\nu)&=\frac{-\pi}{\sin 2 \pi \nu}  \frac{\ep{-zp^2/2 }\, z^{1/2+\nu} p^{1+2\nu} }{\Gamma(1/2-\nu -\xi) \Gamma (1/2+\nu -\xi)} \\
&\sum_{n=0}^\infty\frac{\Gamma(1/2+\nu -\xi+n)}{n! \Gamma(1+2\nu +n)} (z p^2)^n \, . 
\end{split}
\end{equation}
Then the amplitude $A$ is expressed as
\begin{equation}
\frac{- A(z)\sin  2 \pi \nu }{ \pi} =\frac{ z^\nu A_{+\nu}(z)- z^{-\nu} A_{-\nu}(z) }{\Gamma(1/2-\nu -\xi) \Gamma (1/2+\nu -\xi)} \, , 
\label{Anu}
\ee
where 
\be\begin{split}
A_{+\nu}(z)& \doteq \int_{0}^{\infty}d{p} p^{\beta} \ep{- p x\ep{i\pi/4} }\, \ep{-p^2/2 }  \\
&\hspace{1.5cm} \sum_{n=0}^\infty\frac{\Gamma(1/2+\nu -\xi+n)}{n! \Gamma(1+2\nu +n)} (z p^2)^n\\
&=\sum_{n=0}^\infty\frac{\Gamma(1/2+\nu -\xi+n)}{n! \Gamma(1+2\nu +n)} z^n \\
&\hspace{1.5cm}\int_{0}^{\infty}d{p} p^{\beta+2n} \ep{- p x\ep{i\pi/4} }\, \ep{-p^2/2 },
\end{split}
\end{equation}
and where $\beta =\alpha +1+2 \nu$. The last equality is valid only inside the radius of convergence of the power series which is $\abs z < 1/2$. We notice that $z=1$ is not in the radius, this is why we introduced the extra variable $z$. Expanding the oscillating exponential in $x$ as a series, we get
\begin{equation}
\begin{split}
\int_{0}^{\infty}d{p} p^{\beta+2n} &\ep{- p x\ep{i\pi/4} }\, \ep{-p^2/2 } = \sum_{k=0}^\infty \frac{(-1)^k  x^k \ep{i k \pi/4}}{k! }\\
&2^{(\beta -1+k)/2+n} \Gamma(\frac{\beta+1+k+2n}{2}) \, .
\end{split}
\end{equation}
Using this expression, the sum over $n$ can be done and gives 
\begin{equation}
\begin{split}
A_{+\nu}(z)=&\sum_{k=0}^\infty  \frac{(-\sqrt{2i}\,x)^k \Gamma(1/2+\nu -\xi) \Gamma((k+1+\beta)/2)}{k! \Gamma(1+2 \nu) }\\
\times &2^{(\beta-1)/2} \ _2F_1 (\frac12+\nu -\xi,\frac{k+1+\beta}{2};1+2\nu ; 2z )
\end{split}
\end{equation}
Using Eq. (15.3.8) of Ref.~\cite{Abramowitz} and Eq.~\eqref{Anu}, one obtains 
\begin{equation}
\label{Asumk}
\begin{split}
{A(z) } &=\sum_{k=0}^\infty  \frac{\Gamma(1+(k+\alpha)/2+\nu)\Gamma(1+(k+\alpha)/2-\nu)}{ \Gamma(-\xi+{(k+3+\alpha)}/{2})} \\
& \frac{(-\sqrt{2i}\,x)^k }{k! } 2^{\alpha /2}  \ _2F_1 \bigg(\frac12+\nu -\xi,\frac12-\nu -\xi; \\
&\hspace{3cm}-\xi+\frac{k+3+\alpha}{2} ;1-\frac{1}{ 2z} \bigg)\\
\end{split}
\end{equation}

The above expressions are all valid for $\abs z <1/2$. However, since both $A$ and the sum are analytic on $\mathbb{C}$, the result is still valid at $z=1$.

It is then convenient to express the $ _2F_1$ as 
\be
_2F_1 (a,b;c;u) = \sum_{n=0}^\infty \frac{\Gamma(a+n) \Gamma(b+n) \Gamma(c)}{\Gamma(a) \Gamma(b) \Gamma(c+n) n!} u^n\, ,
\ee
 to  split the sum between odd and even $k$, and to notice that each sum over $k$ gives an hypergeometric function
 
\begin{equation}
\label{ampfinal}
\begin{split}
A(1)\! = \!   \sum_{n=0}^\infty & 2^{-n} \frac{\Gamma(1/2+ \nu -\xi +n)\Gamma(1/2- \nu -\xi +n)}{n! \Gamma(1/2+ \nu -\xi )\Gamma(1/2-\nu -\xi) }\! \\
2^{\alpha /2}&\left (\! \! B(\alpha,\frac12) -\ep{i\pi/4} x \sqrt{2} B(\alpha+\frac12,\frac32) \! \right )\! ,
\end{split}
\end{equation}
where
\be
\begin{split}
 &B(\alpha,\epsilon)\! = \!  \Gamma(1+\frac{\alpha}{2}+\nu)\Gamma(1+\frac{\alpha}{2}-\nu)\\
 & \frac{\ _2F_2 (1+\frac{\alpha}{2}+\nu,1+\frac{\alpha}{2}-\nu;\epsilon;n-\xi+\frac{3+\alpha}{2});i\frac{x^2}{2})}{\Gamma(n-\xi+{(3+\alpha)}/{2})}.
 \end{split}
\ee
We verified that the above expression is a solution of the $4^{th}$ order differential Eq.~\eqref{ommodeeq}. From the symmetries of Eq.~\eqref{ommodeeq}, four independent solutions are 
\be
(\phi^{{\rm BD}, U}_{- \om}(X))^*, \, \phi^{{\rm BD}, U}_{+ \om}(X), \, (\phi^{{\rm BD}, U}_{- \om}(-X))^*, \, \phi^{{\rm BD}, U}_{+ \om}(-X) .
\ee
The last two ones give the $V$ modes evaluated at $X$. These four functions are independent because they are orthogonal to each other when using the scalar product of Eq.~\eqref{scalpr}.

In addition, to validate our long calculation, we compared the final expression of Eq.~\eqref{ampfinal} with the original integral of Eq.~\eqref{AMP1} evaluated numerically with Mathematica\textsuperscript{\textregistered}. We found a perfect agreement. 
\vspace{2.4cm}

\bibliographystyle{../../biblio/h-physrev}
\bibliography{../../biblio/bibliopubli}

\end{document}